\documentclass[aps,pra,amsfonts,amssymb,amsmath,twocolumn,showpacs]{revtex4}
\usepackage[dvips]{graphicx,epsfig}
\usepackage{ulem}
\def\ket#1{|#1\rangle}

\def\bracket#1{\langle #1 \rangle}
\def\bracketi#1#2{\langle #1|#2 \rangle}

\def\ketbra#1#2{| #1 \rangle \langle #2 |}
\begin{document}
\title{Photon-arrival detector with a controlled phase flip operation between a photon and a V-type atomic system}
\author{Kunihiro Kojima}
\email{kojima@qci.jst.go.jp}
\author{Akihisa Tomita}
\email{tomita@qci.jst.go.jp}
\affiliation{Quantum Computation and Information Project, ERATO-SORST, JST, Miyukigaoka 34 Tsukuba Ibaraki 305-8501, Japan}
\begin{abstract}
We propose a photon-arrival detector (PAD), which detects the arrival of a signal photon and simultaneously projects the signal input state to a single photon state, with an atom-cavity system. In this proposal, use of a V-type system as the intracavity atom is discussed for implementing the PAD, since V-type systems have been widely studied in the field of solid state, enabling us to miniaturize and integrate that implementation. The performance of the proposed PAD is evaluated for a specific method of the detection process. The proposed PAD is capable of repeating the procedure for detecting the arrival of input photons and it has improves the detection probability so that it has a higher quantum efficiency than those of conventional photodetectors.  
\end{abstract}
\pacs{03.67.Hk, 32.80.-t, 42.50.-p}
\maketitle
\section{Introduction}
The quantum bit (qubit) is a new concept for the basic unit of information. It is an extension of the classical bit and it enables us to develop much more powerful methods for achieving secure communication and information processing. To implement a qubit, the information of a qubit (quantum information) is encoded in the quantum state of a microscopic physical system that is surrounded by a macroscopic system (the environment). Since the quantum state of the microscopic system is sensitive to the environment in general, it is not easy to maintain quantum coherence of the system during the time taken to perform operations on qubits. Thus, quantum information media has to be appropriately chosen in order to maintain quantum coherence during operations on qubits.

Photons and atoms are promising candidates for quantum information media, since their internal states have a high resistance to decoherence due to interaction with the environment. Photons are suited to the transmission and processing of quantum information, and atoms are suited to the storing of that information. Photons and atoms should thus assume complementary roles in quantum information technology. In addition, an efficient interface for transferring quantum information between photons and atoms is required. If quantum information is encoded in the polarization of a photon and in the ground and excited states of an atom, it will be important to generate an atom-photon entangled state. This is because the quantum information encoded in the polarization of a signal photon can be transferred to the internal state of the atom by performing a Bell measurement between a signal photon and a photon entangled with an atom. This information transfer corresponds to one-qubit quantum teleportation by a two-qubit entangled state \cite{bennet}. A controlled-NOT (C-NOT) operation between two qubits is required to implement atom-photon entanglement. This operation changes the bit value of the target qubit to its opposite value only when the bit value of the control qubit is 1. Recently, the implementation and application of the C-NOT operation for photons and atoms have been studied extensively in the field of cavity quantum electrodynamics \cite{duan}, since strong coherent interactions between a single atom and a few photons can be investigated using a small-volume optical cavity \cite{turchette,loock}.

In order to efficiently generate atom-photon entangled states, the C-NOT operation needs to be implemented such that the input photon state must be a single-photon state and the input photon must interact with an atom within a time limited by the decoherence time of the internal state of the atom. For this purpose, it will be useful to confirm the arrival times and the number of the input photons at the input port of the C-NOT gate before commencing the C-NOT operation. Since current single-photon sources \cite{kuhn} are not perfect on-demand sources (i.e., the light that they emit is described as a mixed state of zero- and single-photon) the above-mentioned confirmation will be especially important for efficiently generating atom-photon entangled states.

In this paper, we propose a photon-arrival detector (PAD), which detects the arrival of signal photon and simultaneously projects the signal input state to a single photon state, with an atom-cavity system. The performance of the proposed PAD is quantitatively evaluated by calculating the detection probability of a signal photon. The proposed PAD enables us to repeatedly detect the arrival of the input photons.  The detection probability of the proposed PAD will exceed the quantum efficiency of conventional photodetectors for a small number (i.e., less than ten) attempts, when the linear transmittance per attempt exceeds the quantum efficiency.

\begin{figure}[htbp]
\begin{center} 
\includegraphics[width=4.5cm, angle=0,origin=c]{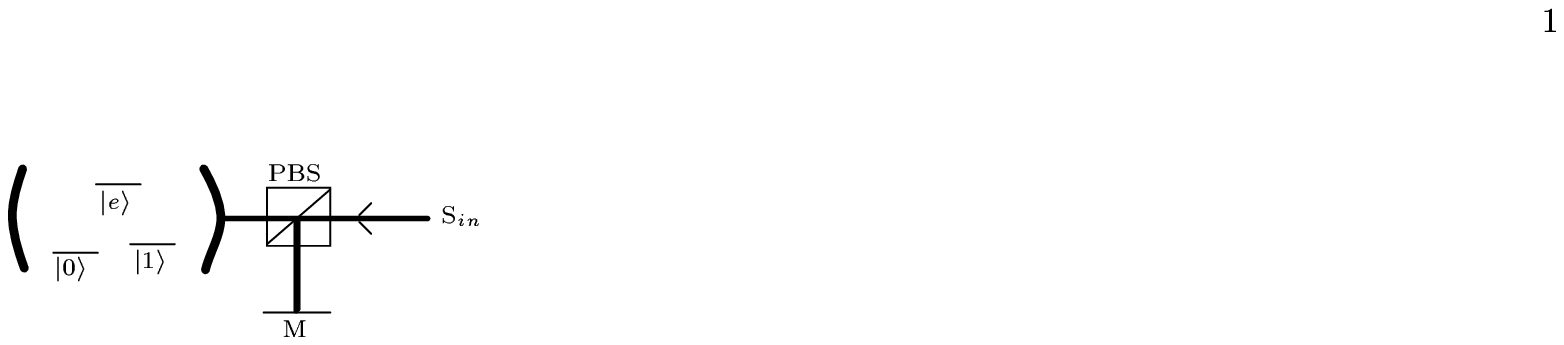}
\caption{\label{fig:cpf} \scriptsize Schematic of the setup to implement CPF between an atom and a photon. With a polarization beam splitter (PBS), the X-polarized  component of the single-photon signal pulse is reflected by the cavity, while the Y-polarized component is reflected by the mirror M. The optical paths from the PBS to the cavity and to the mirror M are assumed to be equal. X and Y are any two mutually orthogonal axes.} \end{center} 
\end{figure}

The key element of the proposed PAD is the implementation of a controlled-phase-flip (CPF) operation by an atom-cavity system. This operation changes the phase of the target qubit by $\pi$ only when the bit values of the control and the target qubits are 1 with each other. Duan and Kimble have proposed a scheme for carrying out CPF \cite{duan}. A schematic setup to implement CPF between an atom and a photon is shown in Fig.~\ref{fig:cpf}. A polarization beam splitter (PBS) transmits the X-polarized component of the signal photon coming from the signal input port ${\rm S}_{in}$ and reflects the Y-polarized component of that photon, where X and Y represent any two mutually orthogonal axes.  The cavity is a one-sided cavity and the intracavity atomic system consists of one excited state $\ket{e}$ and two ground states $\ket{\rm 0}$ and $\ket{\rm 1}$, making it a $\lambda$-type system. The transition between the ground state $\ket{\rm 1}$ and the excited state $\ket{e}$ is resonant with the X-polarized cavity mode. The X-polarized component of the signal photon excites that cavity mode. We evaluated the fidelity of the proposed scheme using the CPF operation described by the unitary operation $U^{CPF}=e^{i \pi \ketbra{0}{0} \otimes \ketbra{\rm X}{\rm X}}$ by numerically calculating the shape mismatch between the input and the output photon pulses. The resultant gate fidelity was more than 99.9 \% in the strong-coupling regime described by $g \gg \kappa$, where $g$ is the coupling strength between the cavity mode and the intracavity atomic system and $\kappa$ is the cavity decay rate.

When processing many photonic or atomic qubits on an individual basis in a laboratory, it is important to miniaturize and integrate devices used to process qubits, and not just to increase the repetition rate of individual devices. In order to miniaturize and integrate CPF implementation in an atom-cavity system and to obtain stable optical responses of an atom-cavity system to single photons, it is important to realize solid-state atom-cavity systems. Solid-state three-level systems have been extensively studied in experiments involving V-type systems, which consist of a single ground state and two excited states \cite{Wu,kosaka-crest}. It is thus interesting to discuss quantitatively CPF in a V-type system together with its potential applications.

The remainder of this paper is organized as follows. In Sec.~II, the scheme for the proposed PAD is explained. In Sec.~III, the Hamiltonian is presented for an atom in a one-sided cavity coupled with input and output radiative fields. The temporal evolution of the intracavity atomic system is then derived for a coherent pulse input (Sec.~IV). The fidelity of the signal output from the atom-cavity system for an ideal output after applying the unitary operation $U_{\rm CPF}$ is estimated by analyzing the responses of the atom-cavity for single-photon input in Sec.~V. The performance of the PAD is then estimated in Sec.~VI, based on the results of Sec.~IV and V. In Sec.~VII, experimental realization of an atom-cavity for the PAD is discussed. In Sec.~VIII, our scheme for CPF is compared with that of Duan and Kimble.

\section{Scheme for photon-arrival detector}

\begin{figure}[htbp]
\begin{center} 
\includegraphics[width=4.0cm, angle=0,origin=c]{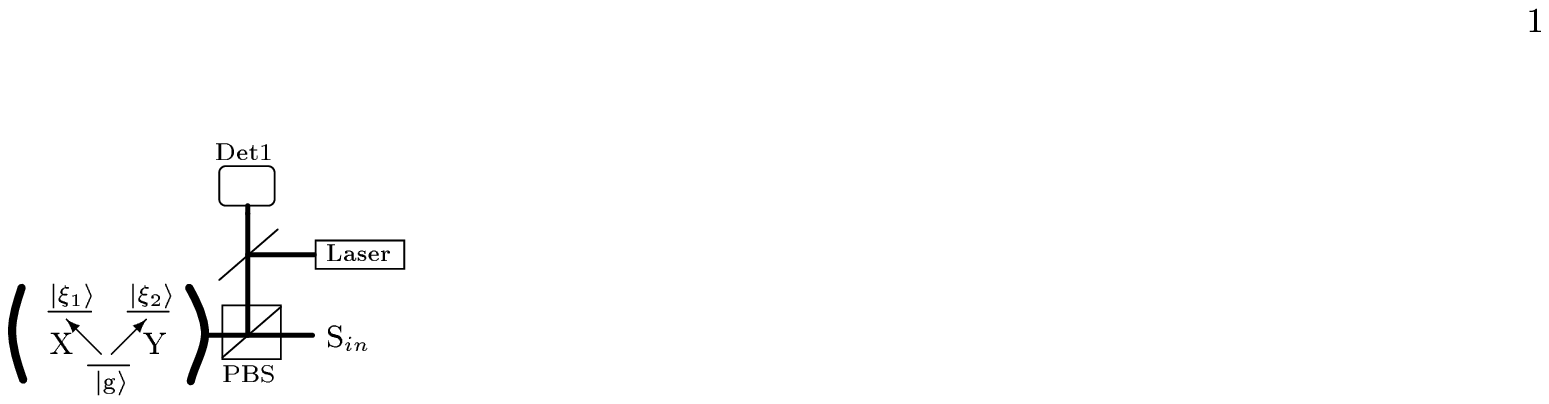}
\caption{\label{fig:qndscheme} \scriptsize Schematic representation of photon-arrival detector.}
\end{center} 
\end{figure}

Figure~\ref{fig:qndscheme} shows the scheme for a PAD based on CPF with a V-type system. The polarization of the signal photon is aligned in the X direction at the input port $S_{in}$. The probability amplitudes of the signal photon form a wavepacket that is directed to the one-sided cavity and then interacts with the atom-cavity system. It is assumed that there are only two allowed modes in the cavity: one is a X-polarized mode and the other is a Y-polarized mode. The X-polarized mode resonantly couples with the transition from the ground state $\ket{\rm g}$ to excited state $\ket{\xi_{1}}$ in the V-type three-level system shown in Fig.~\ref{fig:qndscheme}. Likewise, the Y-polarized mode resonantly couples with the transition to excited state $\ket{\xi_{2}}$. The coupling constants are $g_{1}$ and $g_{2}$, respectively. Such a level structure is realizable by using real atoms \cite{turchette} or quantum dot molecules \cite{tabic} with g-factor engineering \cite{kosaka}.

The procedure for the PAD is described by the unitary operation $U_{\rm PAD}=R \left(-\pi/2 \right) \cdot U_{\rm CPF} \cdot R \left(\pi/2 \right)$, where $R(\theta)\ket{\rm g}=\cos\left(\theta/2\right) \ket{\rm g} + \sin\left(\theta/2\right)\ket{\xi_{2}}$, and $R(\theta) \ket{\xi_{2}} = - \sin\left(\theta/2\right) \ket{\rm g} + \cos \left(\theta/2\right) \ket{\xi_{2}}$. The laser generates pulses to rotate the quantum state of the atomic system between the ground state $\ket{\rm g}$ and the excited state $\ket{\xi_{2}}$. In this procedure, the initial atomic state is the ground state $\ket{\rm g}$. The unitary operation $U_{\rm PAD}$ converts the initial state $\ket{\rm X}_{s} \otimes \ket{\rm g}$ ($\ket{\rm 0}_{s} \otimes \ket{\rm g}$) to the final state $\ket{\rm X}_{s} \otimes \ket{\xi_{2}}$ ($\ket{\rm 0}_{s} \otimes \ket{\rm g}$), where state $\ket{\rm 0}_{s}$ means that there is no photon in the signal input port. The atomic system is changed into the excited state $\ket{\xi_{2}}$ only when there is a signal photon in the signal input port, and it then emits a single Y-polarized photon. The arrival of a signal photon can thus be determined when a Y-polarized photon is detected at the detector $\rm Det1$ and a signal photon is simultaneously detected at the input port ${\rm S}_{in}$. Since this procedure does not affect the polarization state of the signal photon, it can be repeated until a photon is detected at detector $\rm Det1$.  This ability to perform repeated attempts will be useful when detector $\rm Det1$ has a low quantum efficiency. In our calculations, the quantum efficiency (${\rm P}_{\rm eff}$) was assumed to be $10^{-1}$, which is a typical quantum efficiency for detectors at optical communication wavelengths (e.~g.,~InGaAs avalanche photodiode detectors).

It should be noted that when $-\pi/2$ rotation is applied by the laser pulse, the corresponding cavity mode is populated and subsequently emits light having the same polarization as the Y-polarized photon detected by detector $\rm Det1$. We therefore have to wait for the start of the detection of that photon at $\rm Det1$ until the emission from the populated cavity mode is completed.
The desired responses of the atom-cavity system to a signal photon wavepacket for CPF are as follows. When the atomic system is initially in the ground state $\ket{\rm g}$, the phase of the wavepacket is shifted by $\pi$ and the atomic state is unchanged. On the other hand, when the atomic state is initially in one of the two excited states $\ket{\xi_{1}}$ and $\ket{\xi_{2}}$, the phases of both the wavepacket and the atomic state are unchanged. This response is described by the unitary operation $U_{\rm CPF} = \exp \left[-i \pi \ketbra{\rm g}{\rm g} \otimes \ketbra{\rm X}{\rm X}_{s}\right]$, where $\ketbra{\rm X}{\rm X}_{s}$ is a operator on the polarization state of the signal photon.

The performance of the proposed PAD is quantitatively evaluated by calculating the probability that a photon is detected by detector ${\rm Det1}$ in sufficient time after the Y-polarized photon arrives at the detector, and the probability that the output signal state is a single-photon state after detection. We term this latter probability 'credibility'. Note that the proposed PAD includes an atomic system so that its accuracy may be affected by spontaneous emission. It is thus necessary to analyze the effects of spontaneous emission on the scheme before evaluating the performance of the PAD.

In our scheme, there are two spontaneous emissions via non-cavity modes. One is the transition from the excited state \(\xi_{1}\). The other is the transition from the excited state \(\xi_{2}\). The effects of spontaneous emission of the excited state \(\xi_{1}\) on the atom-photon CPF have been quantitatively discussed by Duan and Kimble \cite{duan}. If the spontaneous emission rate $2 \gamma_{1}$ via non-cavity modes is less than the cavity decay rate \(\kappa\), the effects on CPF are negligible in a strong-coupling regime for an input photon with an appropriate duration. In the present model, the effects of spontaneous emission will be negligible since we assume that $\kappa \gg \gamma_{1}$ and that the pulse duration is much longer than $1/\kappa$ in the regime $g_{1} \ge \kappa$.

On the other hand, spontaneous emission from the excited state \(\xi_{2}\) is a new error source introduced by our scheme. The spontaneous emission rate $2 \gamma_{2}$ is given by $2 \gamma_{1} g_{2}/g_{1}$. Note that coupling of the transition to the excited state \(\xi_{2}\) with the Y-polarized mode of the cavity is required to be in a leaky-cavity regime: $\kappa \gg g_{2}$ for our scheme (as explained in Sec.~IV) and the corresponding radiative decay rate through the cavity mode is approximately $\Gamma_{2}=g_{2}^{2}/\kappa$. If the rate $\gamma_2$ is comparable to the radiative decay rate $\Gamma_{2}$, the effects of spontaneous emission decay with the rate $\gamma_{2}$ on the atom-photon CPF will not be negligible. This problem is discussed in Sec.~VII.
\section{Model}
To analyze the responses of the atom in a one-sided cavity to a single-photon pulse input, and the temporal evolution of the atomic state for a laser-pulse input, it is necessary to develop a model of spatiotemporal propagation to and from the atom in a one-sided cavity.

\begin{figure}[ht] 
\begin{center}
\includegraphics[width=5cm]{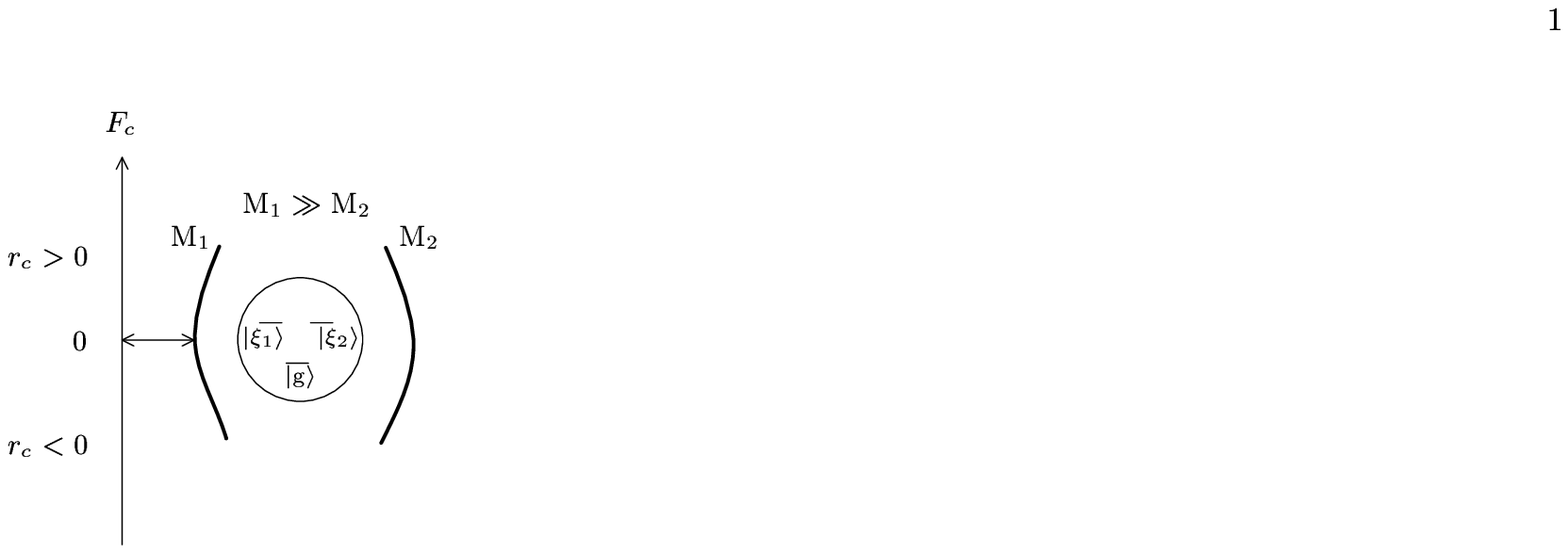}
\caption{\label{fig:model} \scriptsize Schematic representation of cavity geometry.}
\end{center} 
\end{figure}

The proposed model is illustrated in Fig.~\ref{fig:model}. The cavity couples with the radiative field mode \(F_{c}\) via a cavity mirror whose transmittance is ${\rm M}_{1}$ ($\gg {\rm M}_{2}$). In this figure, $\ket{{\rm g}}$, $\ket{\xi_{1}}$ and $\ket{\xi_{2}}$ denote the ground and excited states of the V-type three-level system. It is assumed that there are only two allowed cavity modes. One is the X-polarized mode and the other is the Y-polarized mode. The vertical arrow (\(F_{c}\)) on the left side of the cavity represents the input (\(r_{c} < 0\)) and output (\(r_{c} > 0\)) fields at the cavity, where \(r_{c}\) corresponds to the spatial coordinate. The double-headed arrow at the origin of the vertical arrow represents the coupling between the atom-cavity system and the radiative field \(F_{c}\).

The total Hamiltonian for this model is as follows.
\begin{eqnarray}
&& {} \hat{H} = \sum_{j=1,2} \left( \hat{H}^{(j)}_{F_{c}}+\hat{H}^{(j)}_{int F_{c}}+ \hat{H}^{(j)}_{int ac} \right) \label{eq:exthamiltonian}\\
\mbox{with} && {} \hat{H}^{(j)}_{F_{c}} = \int^{\infty}_{-\infty}dk\ \hbar c k \hat{b}_{F_{cj}}^{\dagger}(k)\hat{b}_{F_{cj}}(k) \nonumber \\
&& {} \hat{H}^{(j)}_{int F_{c}} = \int^{\infty}_{-\infty}dk\ i\hbar\sqrt{\frac{c \kappa}{\pi}} \left( \hat{b}_{F_{cj}}^{\dagger}(k) \hat{a}_{j}-\hat{a}_{j}^{\dagger}\hat{b}_{F_{cj}}(k) \right) \nonumber\\
&& {} \hat{H}^{(j)}_{int ac} = \hbar g_{j} \left( \hat{a}_{j}^{\dagger} \hat{\sigma}^{(j)}_{-}+\hat{\sigma}_{-}^{\dagger (j)}\hat{a}_{j} \right) \nonumber
\end{eqnarray}
where \(\hat{\sigma}^{(j)}_{-} = \ketbra{{\rm g}}{\xi_{j}}\), and \(\hat{a}_{j}\) and \(\hat{b}_{F_{j}}(k)\) are the annihilation operators for the $j$th mode of the cavity and the radiative field \(F_{j}\) (\(j=1,2\)), respectively.

The cavity modes $j=1$ (X-polarized mode) and $2$ (Y-polarized mode) are resonantly coupled to the transitions to the excited states $\ket{\xi_{1}}$ and $\ket{\xi_{2}}$, respectively. All the Hamiltonians presented in this paper have been formulated in a rotating frame defined by the transition frequency of the atomic system \(\omega_{\xi_{1}} = \omega_{\xi_{2}} = \omega_{0}\). The wave vector is also defined in the rotating frame, that is, \(k_{F_{cj}}\) is defined relative to the resonant wave vector \(\omega_{0}/c\). The factor \(\sqrt{c\kappa/\pi}\) is the coupling constant between the cavity modes and the radiative fields, where \(\kappa\) is the cavity decay rate due solely to the coupling of the cavity mode with the radiative field \(F_{cj}\). The factor \(g_{j}\) is the coupling constant between the cavity mode and the atomic system.

\section{Conditions for controlling the intracavity atomic state}
The proposed PAD includes $\pi/2$ and $-\pi/2$ rotations between the ground state $\ket{\rm g}$ and the excited state $\ket{\xi_{2}}$. These rotations are Rabi rotations with Y-polarized laser pulses through the Y-polarized mode of the cavity. In this section, we discuss the optimal conditions of the cavity decay rate $\kappa$, the coupling constant $g_{2}$, and the input laser pulses for preventing entanglement between the Y-polarized cavity mode driven by the laser pulses and the atomic system.

The quantum state of the Y-polarized cavity mode driven by the laser pulse should reach a steady coherent state before interacting with the atomic system to avoid entanglement. An interaction time with the laser pulse comparable to the cavity decay time \(1/\kappa\) is required to reach a steady coherent state, and the effects of the interaction of the atomic system with the intracavity photons become significant for interaction times larger than \(1/(\sqrt{\bar{n}_{a}} g_{2})\), where \(\bar{n}_{a}\) is the average intracavity photon number. The inequality relationship between these two interaction times should therefore be in the leaky-cavity regime \(\kappa \gg \sqrt{\bar{n}_{a}} g_{2}\) to strongly suppress entanglement. This situation is described by solving the Heisenberg equation of motion with the Hamiltonian given by eq.~(\ref{eq:exthamiltonian}) for $j=2$. The temporal evolution of \(\hat{b}_{F_{c2}}(k)\) is then described as
\begin{eqnarray}
\hat{b}_{F_{c2}}(k;t) && {} =e^{-ick(t-t_{i})}\hat{b}_{F_{c2}}(k;t_{i}) \nonumber \\
&& {} \ \ \ \ +\sqrt{\frac{c \kappa}{\pi}} \int^{t}_{t_{i}} dt^{'} \ e^{-ick(t-t^{'})} \hat{a}_{2}(t^{'}) \nonumber \\
\label{eq:fieldc2}
\end{eqnarray}
, where \(t_{i}\) is the initial time of the evolution. The Heisenberg equation of motion for \(\hat{a}_{2}\)
\begin{eqnarray}
\frac{d}{dt} \hat{a}_{2}(t) && {} = -ig_{2} \hat{\sigma}^{(2)}_{-}(t) -\sqrt{\frac{c\kappa}{\pi}}\int^{\infty}_{-\infty} dk \ \hat{b}_{F_{c2}}(k;t) \nonumber \\\label{eq:afield}
\end{eqnarray}
is obtained by substituting eq.~(\ref{eq:fieldc2}) into eq.~(\ref{eq:afield}) as
\begin{eqnarray}
&& {} = -ig_{2} \hat{\sigma}^{(2)}_{-}(t)-\kappa \hat{a}_{2}(t) -\sqrt{2 \kappa c} \hat{b}_{F_{c2}}(-c(t-t_{i});t_{i}) \nonumber \\
\label{eq:transversaleq}
\end{eqnarray}
By integrating eq.~(\ref{eq:transversaleq}), the temporal evolution of \(\hat{a}_{2}\) is obtained as
\begin{eqnarray}
\hat{a}_{2}(t) && {} = -i g_{2} \int^{t}_{t_{i}}dt^{'}\ e^{-\kappa (t-t^{'})} \hat{\sigma}^{(2)}_{-}(t^{'}) + e^{-\kappa(t-t_{i})} \hat{a}_{2}(t_{i}) \nonumber \\
&& {} -\sqrt{2\kappa c}\int^{t}_{t_{i}} dt^{'} \ e^{-\kappa(t-t^{'})} \hat{b}_{F_{c2}}(-c(t^{'}-t_{i});t_{i}) \label{eq:cavityamplitude}
\end{eqnarray}
In the leaky-cavity regime \(\kappa \gg \sqrt{\bar{n}_{a}} g_{2}\), eq.~(\ref{eq:cavityamplitude}) is approximated as
\begin{eqnarray}
&& {} \hat{a}_{2}(t) \simeq -i \frac{g_{2}}{\kappa} \hat{\sigma}^{(2)}_{-}(t) + e^{-\kappa(t-t_{i})} \hat{a}_{2}(t_{i}) \nonumber \\
&& {} -\sqrt{2\kappa c}\int^{t}_{t_{i}} dt^{'} \ e^{-\kappa(t-t^{'})} \hat{b}_{F_{c2}}(-c(t^{'}-t_{i});t_{i}) \label{eq:effectiveamplitude}
\end{eqnarray}
Likewise, the Heisenberg equations of motion for \(\hat{\sigma}^{(2)}_{-}\) and \(\hat{\sigma}^{(2)}_{3}\) are described as
\begin{eqnarray}
\frac{d}{dt} \hat{\sigma}^{(2)}_{-}(t) = ig_{2} \hat{\sigma}^{(2)}_{3}(t) \hat{a}_{2}(t) -\gamma_{2} \hat{\sigma}^{(2)}_{-}(t) \label{eq:sigmaminus}
\end{eqnarray}
\begin{eqnarray}
\frac{d}{dt} \hat{\sigma}^{(2)}_{3}(t) &=& 2ig_{2} \left( \hat{a}_{2}^{\dagger}(t)\hat{\sigma}^{(2)}_{-}(t)-\hat{\sigma}^{(2) \dagger}_{-}(t) \hat{a}_{2}(t) \right) \nonumber \\
&& {} - \gamma_{2} \left( \hat{\sigma}^{(2)}_{3}(t) + {\rm I} \right),
\label{eq:sigmathree} \\
\mbox{where } \hat{\sigma}^{(2)}_{3} &=& \left(\ketbra{\xi_{2}}{\xi_{2}}-\ketbra{\rm g}{\rm g}\right) \nonumber
\end{eqnarray}
The effects of spontaneous emission with a rate of $2\gamma_{2}$ are included in eqs.~(\ref{eq:sigmaminus}) and (\ref{eq:sigmathree}).

Since, in the present analysis, the atom-cavity system is in the ground state prior to the arrival of the coherent laser input pulse, the cavity-state amplitude and the excited-state amplitude are initially zero. Under these assumptions, the initial state is described by
\begin{eqnarray}
&& {} \ket{\Psi(t_{i})} = \ket{\alpha}_{F_{c2}} \otimes \ket{Vac}_{a_{2}} \otimes \ket{\rm g} \mbox{, where } \ket{\alpha} = \hat{D}(\alpha) \ket{Vac} \nonumber\\
&& {} \mbox{with } \hat{D}(\alpha) \equiv \exp \left[\int^{\infty}_{-\infty} dr\ \alpha(r;t_{i})\hat{b}^{\dagger}_{F_{c2}}(r;t_{i})\right. \nonumber \\
&& {} \ \ \ \ \ \ \ \ \ \ \ \ \ \ \ \ \ \ \ \ \ \ \ \ \ \ \ \ \ \ \ \ \ \ \ \ \ \ \ \ \ \ \ \ \ \left.-\alpha^{*}(r;t_{i})\hat{b}_{F_{c2}}(r;t_{i})\right] \nonumber \\
\label{eq:initialstate}
\end{eqnarray}
In eq.~(\ref{eq:initialstate}), \(\alpha(r;t_{i})\) is the initial spatial distribution of the input laser pulse amplitude and $\hat{D}(\alpha)$ is the corresponding displacement operator.

Approximate equations for the time evolution of the longitudinal and transversal components of the atomic system \(\bracket{\hat{\sigma}^{(2)}_{3}(t)}\) and \(\bracket{\hat{\sigma}^{(2)}_{-}(t)}\) under the initial state given by eq.(\ref{eq:initialstate}) are thus obtained by substituting eq.~(\ref{eq:effectiveamplitude}) into eqs.~(\ref{eq:sigmaminus}) and (\ref{eq:sigmathree}) as
\begin{eqnarray}
&& {} \frac{d}{dt} \bracket{\hat{\sigma}^{(2)}_{-}(t)} \simeq -\left(\frac{g_{2}^{2}}{\kappa} + \gamma_{2} \right) \bracket{\hat{\sigma}^{(2)}_{-}(t)} \nonumber \\
&& {} \ \ \ \ \ \ +ig_{2}\sqrt{2\kappa} \bracket{\hat{\sigma}^{(2)}_{3}(t)} \int^{t}_{t_{i}}dt^{'} \ e^{-\kappa(t-t^{'})} \epsilon_{in}(t^{'}) \nonumber \\
 \label{eq:transversal} \\
&& {} \frac{d}{dt} \bracket{\hat{\sigma}^{(2)}_{3}(t)} \simeq -2 \left(\frac{g_{2}^{2}}{\kappa} + \gamma_{2} \right) \left(\bracket{\hat{\sigma}^{(2)}_{3}(t)} + 1\right) +2ig_{2} \sqrt{2 \kappa} \nonumber \\
&& {} \times \int^{t}_{t_{i}}dt^{'} \ e^{-\kappa(t-t^{'})} \left(\epsilon_{in}^{*} (t^{'})\bracket{\hat{\sigma}^{(2)}_{-}(t)}-\bracket{\hat{\sigma}^{(2)}_{-}(t)}^{*}\epsilon_{in} (t^{'})\right), \nonumber \\
\label{eq:longitudinal}
\end{eqnarray}
where \(\epsilon_{in}(t) = \sqrt{c}\alpha(-c(t-t_{i});t_{i})\).
\begin{eqnarray}
\epsilon_{in}(t) =
\begin{cases}
0 & \text{for $t \leq t_{i}$} \\
e^{i \phi}\epsilon & \text{for $t_{i} \leq t \leq t_{f}$} \\
0 & \text{for $t_{f} \leq t$}
\end{cases}
\end{eqnarray}
The solutions for eqs.~(\ref{eq:transversal}) and (\ref{eq:longitudinal}) are given by
\begin{eqnarray}
&& {} \bracket{\hat{\sigma}^{(2)}_{-}(t)} \nonumber \\
&& {} \ \ \ =
\begin{cases}
\bracket{\hat{\sigma}^{(2)}_{-}(t)} & \text{for $t \leq t_{i}$} \\
v_{-}(t,\phi) + s_{-} & \text{for $t_{i} \leq t \leq t_{f}$} \\
\bracket{\hat{\sigma}^{(2)}_{-}(t_{f})} e^{-\left(\Gamma_{2}+\gamma_{2}\right)(t-t_{f})} & \text{for $t_{f} < t$}
\end{cases}, \nonumber \\
\end{eqnarray}
where 
\begin{eqnarray}
&& {} v_{-}(t,\phi) = \frac{1}{8 \sqrt{2 \beta_{2} \bar{n}_{in}}} \left( (1-\Upsilon)e^{\lambda_{+}t}(f_{+} + I_{+}(\phi))\right.\nonumber \\
&& {} \ \ \ \ \ \ \ \ \ \ \ \ \ \ \ \ \ \ \ \ \ \ \ \left. + (1+\Upsilon)e^{\lambda_{-}t}(f_{-} + I_{-}(\phi)) \right) \nonumber \\
&& {} s_{-} =   \frac{1}{8 \sqrt{2 \beta_{2} \bar{n}_{in}}} ((1-\Upsilon)f_{+}+(1+\Upsilon)f_{-}) \nonumber
\end{eqnarray}
with $\Gamma_{2}=g^{2}_{2}/\kappa$, $\beta_{2}=\Gamma_{2}/\left(\Gamma_{2}+\gamma_{2}\right)$, $\bar{n}_{in}=\epsilon^{2}/\left(\Gamma_{2}+\gamma_{2}\right)$, $\Upsilon=\sqrt{1-32 \beta_{2} \bar{n}_{in}}$,
\begin{eqnarray}
&& {} \lambda_{\pm} = -\frac{\gamma_{2}+\Gamma_{2}}{2}\left(3 \pm \Upsilon \right), f_{\pm} = \pm \frac{16 \bar{n}_{in}}{\Upsilon (3 \pm \Upsilon)} \mbox{, and} \nonumber \\
&& {} I_{\pm} (\phi) = \frac{\pm 1}{\Upsilon} \left(\frac{1 \pm \Upsilon}{2}(\bracket{\hat{\sigma}^{(2)}_{3}(t_{i})}+1)\right. \nonumber \\
&& {} \ \ \ \ \ \ \ \ \ \ \ \ \ \ \ \ \ \left. -4\sqrt{2 \beta_{2} \bar{n}_{in}} e^{-i \phi} \bracket{\hat{\sigma}^{(2)}_{-}(t_{i})}\right). \nonumber \\
 \label{eq:semiclassica}
\end{eqnarray}
\begin{eqnarray}
&& {} \bracket{\hat{\sigma}^{(2)}_{3}(t)} \nonumber \\
&& {} =
\begin{cases}
\bracket{\hat{\sigma}^{(2)}_{3}(t)} & \text{for $t \leq t_{i}$} \\
v_{3}(t,\phi) + s_{3} & \text{for $t_{i} \leq t \leq t_{f}$} \\
\left(\bracket{\hat{\sigma}^{(2)}_{3}(t_{f})} + 1\right)e^{-2 \left(\Gamma_{2}+\gamma_{2}\right)(t-t_{f})} -1& \text{for $t_{f} < t$}
\end{cases}, \nonumber
\end{eqnarray}
\begin{eqnarray}
&& {}
\end{eqnarray}
where
\begin{eqnarray}
&& {} v_{3}(t,\phi) = e^{\lambda_{+}(t-t_{i})} \left(I_{+}(\phi) + f_{+}\right) + e^{\lambda_{-}(t-t_{i})} \left(I_{-}(\phi) + f_{-}\right) \nonumber \\
&& {} s_{3}    = -\left(f_{+} + f_{-} + 1 \right). \nonumber
\end{eqnarray}
\(s_{-}\) and \(s_{3}\) are steady states that become dominant in the long-pulse limit. Rabi oscillation in time-dependent components \(v_{-}(t,\phi)\) and \(v_{3}(t,\phi)\) becomes dominant when the average input photon number \(\bar{n}_{in}\) given in eq.~(\ref{eq:semiclassica}) exceeds \(1/(32\beta_{2})\). In the following discussion, we treat the case when $\gamma_{2}=0$ in order to investigate the effects of radiative relaxation through the cavity mode. The effects of spontaneous emission on the atom-photon CPF are discussed in Sec.~VII.

The transversal and longitudinal components relax with a relaxation rate of \(\Gamma_{2}\) during the interaction of the laser pulse with the signal photons. This relaxation suppresses the preparation of the initial atomic state (a half-way state) $\left(\ket{\rm g} + \ket{\xi_{2}}\right)/\sqrt{2} \equiv \ket{\Phi_{h}}$ for $\phi=-\pi/2$ and this will reduce the probability of detecting a signal photon in the PAD. To investigate the influence of dipole relaxation on that preparation, the Bloch vector representation is introduced below.

The Bloch vector representation for the atomic state $\rho_{atom}(t) = \frac{1}{2} \left(\hat{{\rm I}} + \sum^{3}_{i=1}P_{i}(t) \hat{\sigma}^{(2)}_{i}\right)$ is described using the transverse and longitudinal components given by
\begin{eqnarray}
&& {} P_{1}(t) = \bracket{\hat{\sigma}^{(2)}_{-}(t)} + \bracket{\hat{\sigma}^{(2)}_{-}(t)}^{*} \nonumber \\
&& {} P_{2}(t) = -i \left(\bracket{\hat{\sigma}^{(2)}_{-}(t)} - \bracket{\hat{\sigma}^{(2)}_{-}(t)}^{*}\right) \nonumber \\ 
&& {} \mbox{, and } P_{3}(t) = \bracket{\hat{\sigma}^{(2)}_{3}(t)}. \nonumber \\
\label{eq:bloch}
\end{eqnarray}

The density matrix $\rho_{atom}(t)$ is rewritten by using the half-way state $\ket{\Phi_{h}}$ for the phase of the input laser $\phi=-\pi/2$ as
\begin{eqnarray}
\rho_{atom}(t) && = {\rm P}_{h}(t) \ketbra{\Phi_{h}}{\Phi_{h}} \nonumber \\
&& {} + {\rm P}^{\perp}_{h}(t) \ketbra{\Phi^{\perp}_{h}}{\Phi^{\perp}_{h}} \nonumber \\
&& {} + \frac{P_{3}(t)}{2} \left(\ketbra{\Phi_{h}}{\Phi^{\perp}_{h}} + \ketbra{\Phi^{\perp}_{h}}{\Phi_{h}} \right) \label{eq:halfrepresentation}, \\
\mbox{ where} && {} {\rm P}_{h}(t) \equiv \frac{1+P_{1}(t)}{2} \mbox{ and } {\rm P}^{\perp}_{h}(t) \equiv \frac{1-P_{1}(t)}{2}. \nonumber
\end{eqnarray}
The probability ${\rm P}^{\perp}_{h}(t)$ given in eq.~(\ref{eq:halfrepresentation}) is the projection probability for the orthogonal half-way state $\ket{\Phi^{\perp}_{h}}$. This probability is increased by dipole relaxation. The state $\ket{\Phi^{\perp}_{h}}$ changes to the excited state $\ket{\xi_{2}}$ after the second rotation ${\rm R}(-\pi/2)$ in the procedure of the PAD when there is no photon in the signal input port. This reduces the accuracy of the proposed PAD. The PAD therefore requires that the probability ${\rm P}^{\perp}_{h}(t)$ be negligibly small at the stage of the first rotation ${\rm R}(\pi/2)$ and during the interaction of the atom-cavity system with the signal input. For this reason, it is important to increase the Rabi frequency to ensure that rotation finishes before the effects of radiative relaxation become significant. However, it is not possible to increase the average input photon number $\bar{n}_{in}$ arbitrarily to increase the Rabi rotation frequency, due to the restriction to be in the leaky-cavity regime \(\kappa \gg \sqrt{\bar{n}_{a}} g_{2}\). To achieve a projection probability ${\rm P}_{h}(t)$ of more than $0.99$, we analyzed the time evolution of ${\rm P}_{h}(t)$.

\begin{figure}[htbp]
\begin{center} 
\includegraphics[width=5.0cm]{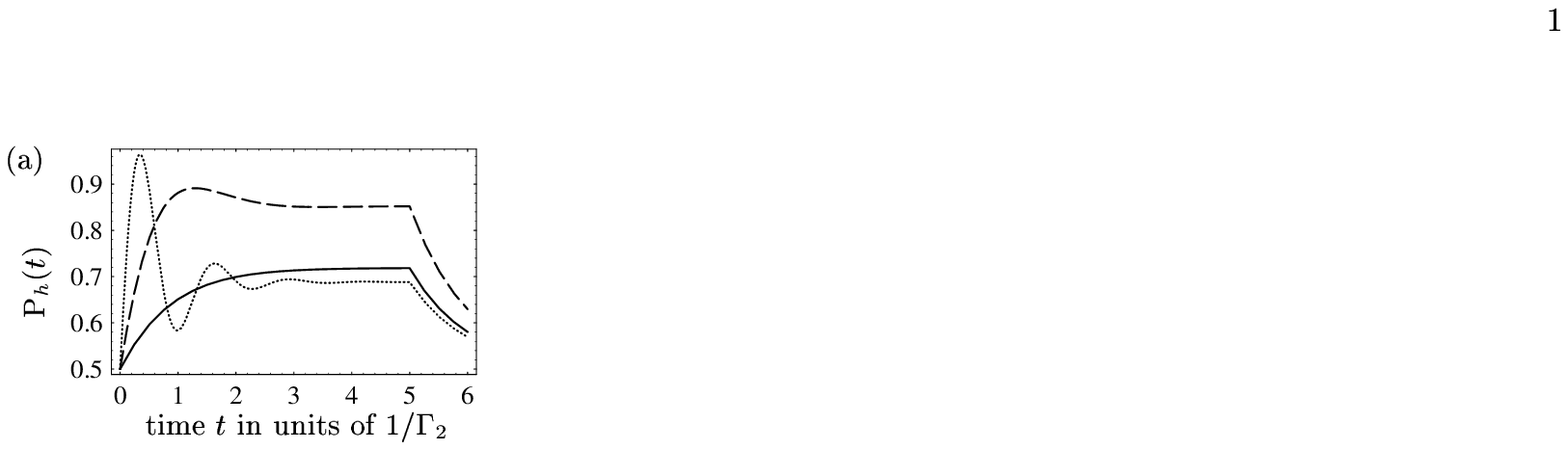}\\
\vspace{0.5cm}
\includegraphics[width=5.0cm]{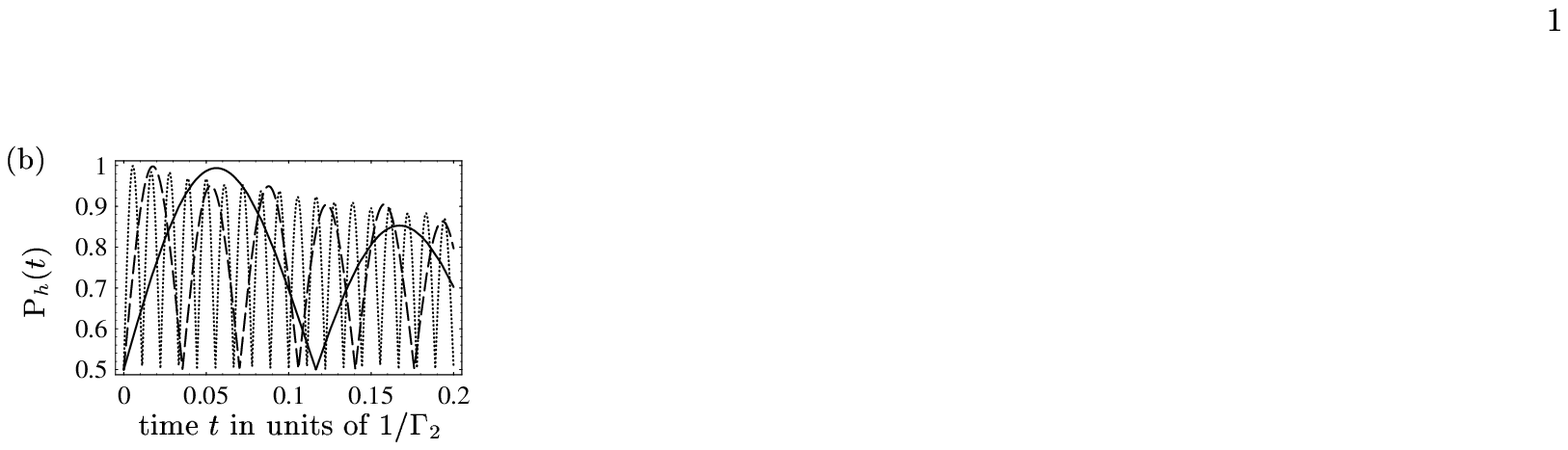}
\caption{\label{fig:halfway} \scriptsize Projection probabilities (a) ${\rm P}_{h}(t)$ for weak coherent pulse inputs of duration $5/\Gamma_{2}$ and average input photon numbers of $\bar{n}_{in}=$ $0.03$ (solid line), $0.3$ (broken line), and $3$ (dotted line) starting from the ground state $\ket{\rm g}$, and (b) for strong coherent pulse inputs of duration $5/\Gamma_{2}$ and average input photon numbers of $\bar{n}_{in}=$ $10^{2}$ (solid line), $10^{3}$ (broken line), and $10^{4}$ (dotted line).}
\end{center} 
\end{figure}

Figure~\ref{fig:halfway}(a) shows the time evolution of the projection probability to the half-way state for weak coherent input pulses having a duration of $5/\Gamma_{2}$ and average input photon numbers of $\bar{n}_{in}=$ $0.03$ (solid line), $0.3$ (broken line), and $3$ (dotted line) starting from the ground state $\ket{\rm g}$. For $\bar{n}_{in}=0.3$, the effect of Rabi oscillation is slight for short time periods of less than \(2/\Gamma_{2}\), whereas for $\bar{n}_{in}=3$ the  effect of Rabi oscillation become significant in this time region causing the maximum probability to increase up to about $0.96$. The average input photon number should thus be much larger than $\bar{n}_{in}=3$ for preparing the initial atomic state. For example, the maximum probability exceeds $0.99$ when $\bar{n}_{in} > 50$.

Figure~\ref{fig:halfway}(b) shows the time evolution of the projection probability to a half-way state for strong coherent input pulses having durations of $5/\Gamma_{2}$ and an average input photon number of $\bar{n}_{in}=$ $10^{2}$ (solid line), $10^{3}$ (broken line), and $10^{4}$ (dotted line). The maximum probabilities are $0.993$, $0.997$, and $0.999$ for $\bar{n}_{in}=10^{2}$, $10^{3}$, and $10^{4}$, respectively. These results are valid in the leaky-cavity regime \(\kappa \gg g_{2} \sqrt{\bar{n}_{a}}\), where the average intracavity photon number is about \(\bar{n}_{a} \simeq 2 \bar{n}_{in} \cdot \Gamma_{2}/\kappa\). For example, when \(g_{2}\) is assumed to be \(10^{-2} \cdot \kappa\), the average intracavity photon numbers for $\bar{n}_{in}=10^{2}$, $10^{3}$, and $10^{4}$ are \(\bar{n}_{a} \simeq 2 \cdot 10^{-2}\), \(2 \cdot 10^{-1}\), and \(2\), respectively. All cases satisfy the leaky-cavity regime under the above-mentioned assumption: $g_{2}=10^{-2}\kappa$. However, an average input photon number $\bar{n}_{in}$ of larger than $10^{7}$ does not fall in the leaky-cavity regime. This implies that the upper-limit of the average input photon number $\bar{n}_{in}$ depends on the scale of $g_{2}$ due to the need to be in the leaky-cavity regime.

In our PAD, we will have to wait for the start of the detection at $\rm Det1$ until intra cavity photons introduced by laser pulses go away from the devices for the PAD, as mentioned in the second last paragraph of Sec.~II. We evaluate this waiting time using the exponential decay of the average intracavity photon number $\bar{n}_{a}$ for the decay rate $\kappa$, described by $\bar{n}_{a} e^{-2\kappa\tau}$. For example, the times for the photon number to decrease to less than $10^{-3}$ are $\tau = 3/\kappa$, $5.3/\kappa$ and $7.6/\kappa$ for average intracavity photon numbers of \(\bar{n}_{a} \simeq 2 \cdot 10^{-2}\), \(2 \cdot 10^{-1}\), and \(2\), respectively. The probability of the excited state $\ket{\xi_{2}}$ gives the upper limit of the detection probability for the probe photon emitted from the intracavity atom. The upper limit decreases to at least $e^{-2\Gamma_{2} \tau}=0.9994$, $0.9989$, and $0.9985$ for waiting times of $\tau = 3/\kappa$, $5.3/\kappa$, and $7.6/\kappa$ when $\Gamma_{2}=g^{2}_{2}/\kappa=10^{-4}\kappa$.

\section{Responses of the atom-cavity system to single-photon input}
After the first rotation ${\rm R}(\pi/2)$ in the procedure of the PAD, the unitary operator $U_{\rm CPF}$ operates on the X-polarized state of the signal photon and the ground state $\ket{\rm g}$ without changing the pulse shape of the signal photon. Note that, in the present proposal, the unitary operation is assumed to be realized by the interaction of the signal photon with the atom in the one-sided cavity. Such an interaction generally alters the pulse shape of the signal pulse in the cavity output \cite{multimode}. This change increases when the procedure of the PAD is repeatedly applied and reduces the fidelity of the output photon pulse from the atom-cavity system after applying the unitary operator $U_{\rm CPF}$. It will thus be necessary to analyze the fidelity in order to evaluate the performance of repeated applications of the proposed PAD.

When the input state of the signal photon and the intracavity atomic system are described as 
\begin{eqnarray}
&& {} \ket{\Psi_{\rm in}} \equiv \ket{\psi_{in}} \otimes \left(c_{\rm g}\ket{\rm g} + c_{\xi_{2}}\ket{\xi_{2}}\right), \\
&& {} \mbox{ where\ \ \ } \ket{\psi_{in}} \equiv \int^{\infty}_{-\infty} dr \ \psi_{in}(r) \ket{r}, \nonumber
\end{eqnarray}
which is the X-polarized signal photon input state, and the wavefunction $\psi_{in}(r)$ describes the spatiotemporal envelope of that photon pulse on the input light field (the region $r_{c} < 0$ in Fig. 3). The output state of the ideal unitary operation $U_{\rm CPF}$ is described as
\begin{eqnarray}
&& {} \ket{\Psi^{\rm ideal}_{\rm out}} \equiv \ket{\psi^{'}_{in}} \otimes \left(c_{\xi_{2}}\ket{\xi_{2}} - c_{\rm g}\ket{\rm g} \right), \\
&& {} \mbox{ where } \ket{\psi^{'}_{in}} \equiv \int^{\infty}_{-\infty} dr \ \psi_{in}(r+d) \ket{r}. \nonumber
\end{eqnarray}
The output state differs from the input state by the phase between the ground state $\ket{\rm g}$ and the excited state $\ket{\xi_{2}}$. The parameter $d$ represents the relative delay time from the input photon pulse described by the state $\ket{\psi_{in}}$, and is $d=0$ in ideal operation. However, a finite $d$ will not affect the performance of the CPF.

The output state from the atom-cavity system is described by
\begin{eqnarray}
\ket{\Psi}_{\rm out} = \left(c_{\rm g} \ket{\psi^{\rm g}_{\rm out}} \otimes \ket{\rm g} + c_{\xi_{2}} \ket{\psi^{\rm \xi_{2}}_{out}} \otimes \ket{\xi_{2}}\right)
\end{eqnarray}
The states $\ket{\psi^{\rm g}_{\rm out}}$ and $\ket{\psi^{\rm \xi_{2}}_{out}}$ are the output photon states when the atomic system is in the ground and in the excited states, respectively.

The fidelity of the output state to the ideal one is thus given by
\begin{eqnarray}
F(d) \equiv \left|\bracketi{\Psi^{\rm ideal}_{\rm out}}{\Psi_{\rm out}}\right| = \left|c_{\xi_{2}}\right|^{2}F^{\xi_{2}}_{int}(d)+\left|c_{\rm g}\right|^{2}F^{\rm g}_{int}(d), \nonumber \\
\end{eqnarray}
where
\begin{eqnarray}
F^{\xi_{2}}_{int}(d) = \bracketi{\psi^{'}_{in}}{\psi^{\xi_{2}}_{\rm out}} \mbox{, } F^{\rm g}_{int}(d) = -\bracketi{\psi^{'}_{in}}{\psi^{\rm g}_{\rm out}}. \nonumber
\end{eqnarray}
In particular, when $F^{\xi_{2}}_{int}(d)$ is equal to $F^{\rm g}_{int}(d)$ for a certain delay time $d=d_{p}$, the fidelity $F(d)$ is constant for all values of $c_{g}$ and $c_{\xi_{2}}$. We then define the 'gate fidelity' by ${\rm F}_{int}=F^{\xi_{2}}_{int}(d_{p})=F^{\rm g}_{int}(d_{p})$.

To evaluate the gate fidelity ${\rm F}_{int}$, it is necessary to determine the responses of the atom-cavity system to single-photon input when the V-type three-level system is initially in the ground state $\ket{\rm g}$. In this section, we derive its solution and evaluate the gate fidelity ${\rm F}_{int}$ for both strong and weak coupling regimes. The gate fidelity ${\rm F}_{int}$ should be larger than $0.99$ to effectively suppress the influence of entanglement between the atomic system and the output photon on the atomic state in the second rotation ${\rm R}(-\pi/2)$.

The state of the field-atom-cavity system for single-photon processes can be expanded on the basis of the wavenumber state \(\ket{k_{1}}\) of the radiative field, the excited state of the V-type three-level system \(\ket{ {\rm E}_{1}}\), and the cavity single-photon state \(\ket{{\rm C}_{1}}\). The state \(\ket{k_{1}}\) denotes a state with the atom in the ground state {\rm g}, the cavity mode \(a_{1}\) in the vacuum state, and one mode of the "$F_{c1}$ field" \(k_{1}\) in the first excited state, with the remaining states being the vacuum state (i.e., \(\ket{k_{1}}=\ket{{\rm g},0_{a_{1}},1_{k_{1}}}\)). Likewise, \(\ket{{\rm C}_{1}}=\ket{{\rm g},1_{a_{1}},0_{k_{1}}}\), and \(\ket{{\rm E}_{1}}=\ket{\xi_{1},0_{a_{1}},0_{k_{1}}}\). The quantum state for the single-photon process can then be written as
\begin{eqnarray}
\ket{\Psi(t)} && {} = \Phi({\rm E}_{1};t)\ket{{\rm E}_{1}} + \Lambda({\rm C}_{1};t) \ket{{\rm C}_{1}} \nonumber \\
&& {} + \int dk_{1} \ \psi(k_{1};t)\ket{k_{1}} \label{eq:statebasis}
\end{eqnarray}
On the base of this, the Hamiltonian given by eq.~(\ref{eq:exthamiltonian}) for $j=1$ can be expressed as
\begin{eqnarray}
&& {} \hat{H}_{1ph} =\hbar c\hat{k}_{1} \nonumber \\
&& {} + i\hbar\sqrt{\frac{c\kappa}{\pi}} \int^{\infty}_{-\infty} dk_{1} \left(\ketbra{k_{1}}{{\rm C}_{1}}-\ketbra{{\rm C}_{1}}{k_{1}}\right) \nonumber \\
&& {} + \hbar g_{1} \left(\ketbra{{\rm C}_{1}}{{{\rm E}_{1}}}+\ketbra{{\rm E}_{1}}{{\rm C}_{1}}\right) \nonumber \\
&& {} \mbox{where } \hat{k}_{1} = \int^{\infty}_{-\infty} dk_{1} \ \ketbra{k_{1}}{k_{1}}. \label{eq:onephotonhamiltonian}
\end{eqnarray}
The equations for the temporal evolution of the probability amplitudes \(\Phi({\rm E}_{1};t)\), \(\Lambda({\rm C}_{1};t)\), and \(\psi(k_{1};t)\) can thus be obtained from the Schr\"odinger equation \(i \hbar d/dt \ket{\Psi(t)} = \hat{H} \ket{\Psi(t)}\) using eqs.~(\ref{eq:statebasis}) and (\ref{eq:onephotonhamiltonian}) as follows.
\begin{eqnarray}
&& \frac{d}{dt} \Phi({\rm E}_{1};t) = -ig \Lambda ({\rm C}_{1};t) \label{eq:excitedamp}\\
&& \frac{d}{dt} \Lambda({\rm C}_{1};t) = -ig \Phi (\xi_{1};t) - \sqrt{\frac{c \kappa}{\pi}} \int dk_{1} \ \psi(k_{1};t) \nonumber \\
&& \label{eq:cavityamp}\\
&& \frac{d}{dt} \psi(k_{1};t) = -i k_{1}c \psi(k_{1};t) +\sqrt{\frac{c\kappa}{\pi}} \Lambda({\rm C}_{1};t) \label{eq:fieldone}
\end{eqnarray}
The evolution \(\psi(k_{1};t)\) can be obtained by integrating eq.~(\ref{eq:fieldone}):
\begin{eqnarray}
&& {} \psi(k_{1};t) = e^{-i k_{1}c \left(t-t_{i}\right)} \psi(k_{1};t_{i}) \nonumber \\
&& {} +\sqrt{\frac{c\kappa}{\pi}} \int^{t}_{t_{i}}dt^{'}\ e^{-ik_{1}c \left(t-t^{'}\right)} \Lambda({\rm C}_{1};t^{'}) \label{eq:fieldonesol}
\end{eqnarray}
where \(t_{i}\) is the initial time of the evolution.

To describe the evolution in real space, the results taking the Fourier transformation of eq.~(\ref{eq:fieldonesol}) are given by
\begin{eqnarray}
&& {} \psi(r_{1};t) \equiv 
\begin{cases}
\frac{1}{\sqrt{2\pi}} \int^{\infty}_{-\infty}dk_{1} \ e^{ik_{1} \cdot r_{1}}\psi(k_{1};t) \mbox{ for } r_{1} < 0 \\
-\frac{1}{\sqrt{2\pi}} \int^{\infty}_{-\infty}dk_{1} \ e^{ik_{1} \cdot r_{1}}\psi(k_{1};t) \mbox{ for } r_{1} > 0
\end{cases} \nonumber \\
\label{eq:fouriertrans}
\end{eqnarray}

The real-space representation of the temporal evolution on the field \(F_{c1}\) is then given by
\begin{eqnarray}
&& \psi(r_{1};t) \nonumber \\
&& = 
\begin{cases}
\psi(r_{1}-c(t-t_{i});t_{i}) \text{\ for $r_{1} < 0$} \\
-\psi(r_{1}-c(t-t_{i});t_{i}) \text{\ for $c(t-t_{i}) < r_{1}$} \\
-\psi(r_{1}-c(t-t_{i});t_{i}) - \sqrt{\frac{2\kappa}{c}} \Lambda({\rm C}_{1};t-\frac{r_{1}}{c}) \\ \text{\ \ \ \ \ \ \ \ \ \ \ \ \ \ \ \ \ \ \ \ \ \ \ \ \ \ \ \ \ \ \ \ \ for $0 < r_{1} < c(t-t_{i})$.}
\end{cases} \nonumber
\end{eqnarray}
\vspace{-0.5cm}
\begin{eqnarray}
&& \label{eq:realspaceampone}
\end{eqnarray}
The first case corresponds to the amplitude of a single photon propagating on an incoming field \(r_{1} < 0\) and the second case corresponds to the amplitude of a single photon reflected by the cavity mirror and then propagating on the outgoing field \(r_{1} > 0\). The third case consists of two parts: the component reflected by the mirror, and the amplitude of a single photon re-emitted by the outgoing field \(r_{1} > 0\) after being absorbed by the cavity.

The temporal evolution of the cavity single-photon amplitude can be obtained by solving a matrix representation consisting of eqs.~(\ref{eq:excitedamp}) and (\ref{eq:cavityamp}):
\begin{eqnarray}
\frac{d}{dt} \begin{pmatrix}
\Phi({\rm E}_{1};t) \\ \Lambda({\rm C}_{1};t)
\end{pmatrix}
&& {} =-
\begin{pmatrix}
0 & ig \\ ig & \kappa
\end{pmatrix}
\begin{pmatrix}
\Phi({\rm E}_{1};t) \\ \Lambda({\rm C}_{1};t)
\end{pmatrix} \nonumber \\
&& {} -\sqrt{2c\kappa}
\begin{pmatrix}
0 \\
\psi(-c(t-t_{i});t_{i})
\end{pmatrix}
\end{eqnarray}
, and using the Fourier transform (\ref{eq:fouriertrans}). Since, in the present analysis, the atom-cavity system is in the ground state $\ket{\rm g}$ before the arrival of the signal photon, the initial amplitudes of the cavity state \(\Lambda({\rm C}_{1};t_{i})\) and the excited state \(\Phi({\rm E}_{1};t_{i})\) are zero, and the field amplitude \(\psi_{1}(r;t_{i})\) is zero for the region \(r_{1}>0\). Under these conditions, 
\begin{eqnarray}
&& {} \Lambda({\rm C}_{1};t) = \frac{-h_{-}(t)\Omega_{-}+ h_{+}(t)\Omega_{+}}{\Omega_{+} - \Omega_{-}}, \label{eq:simplecavityamp} \\
&& {} \mbox{where } h_{\pm}(t) = -\sqrt{2c\kappa} \int^{t}_{t_{i}}dt^{'} \ e^{-\Omega_{\pm}(t-t^{'})}\psi(-c(t-t^{'});t_{i}), \nonumber \\
&& {} \mbox{and } \Omega_{\pm} = \left(\kappa \pm \sqrt{\kappa^{2}-4g_{1}^{2}}\right)/2 \mbox{ for $g_{1} \neq \kappa/2$.} \nonumber
\end{eqnarray}
When $g_{1}$ is larger than $\kappa/2$, the imaginary component of $\Omega_{\pm}$ corresponds to the vacuum Rabi frequency. For a spontaneous emission rate of $2 \cdot \gamma_{1}$, the rate $\kappa$ in $\Omega_{\pm}$ is substituted by the rate $\gamma_{1}+\kappa$, which corresponds to the half width of the spectrum of the atom-cavity system at the Rabi frequency. The vacuum Rabi splitting suppresses the effects of spontaneous emission on the pulse shape of the signal input. This is because, when the center frequency of the signal pulse is tuned to the transition frequency of the excited state $\ket{\xi_{1}}$ and the pulse duration increases beyond $2/(\kappa+\gamma_{1})$, the overlap between the spectrum of the signal with that of the atom-cavity system at the Rabi frequency decreases significantly. In real space, it corresponds to a decrease in the amplitude of the cavity state $\Lambda({\rm C}_{1};t)$ so that the signal no longer excites the atom-cavity system and is not totally reflected by the cavity mirror.

To investigate the outgoing amplitude \(\psi(r_{1}>0;t)\) for an arbitrary incoming amplitude under the above-mentioned initial conditions, it is convenient to represent the outgoing amplitudes as matrix elements of the evolution operator in the following manner.
\begin{eqnarray}
\psi(r_{1};t) && {} = \bracketi{r_{1}}{\Psi(t)} \nonumber \\
                      && {} = \int^{\infty}_{-\infty}dr^{'}_{1} \  {\rm u}_{1ph}(r_{1},r^{'}_{1};t-t_{i}) \nonumber \\
&& {} \times \psi_{1}(r^{'}_{1};t_{i}) \label{eq:matrixrepresentl}
\end{eqnarray}
Here, \({\rm u}_{1ph}(r_{1},r^{'}_{1};t-t_{i})\) is the matrix element of the evolution operator \(e^{-\frac{i}{\hbar}\hat{H}_{\rm 1 ph}}\), representing the transition probability amplitude from the state \(\ket{r^{'}_{1}}\) at time \(t_{i}\) to the state \(\ket{r_{1}}\) at time \(t\), where \(\ket{r_{1}} \equiv \frac{1}{\sqrt{2\pi}}\int^{\infty}_{-\infty}dk_{1} \ e^{-ik_{1}r_{1}} \ket{k_{1}}\). The output wavefunction describes the far-field state of the photons after they have interacted with the atom-cavity system. In general, the wavefunction of a single photon propagating in space is given by $\psi(r_{1};t)=\psi(r_{1}-ct;t_{i})$. Eq.~(\ref{eq:matrixrepresentl}) can therefore be simplified by transforming to a moving coordinate system, i.~e., $r_{1}-ct=r$. In this coordinate system, the output wavefunction in the outgoing far field is expressed as
\begin{eqnarray}
\psi_{out}(r) = \int^{\infty}_{-\infty}dr^{'} \  {\rm u}_{1ph}(r,r^{'}) \times \psi_{in}(r^{'})
\end{eqnarray}
, where ${\rm u}_{1ph}(r,r^{'})$ is given by
\begin{eqnarray}
&& {} {\rm u}_{1ph}(r;r^{'}) = {\rm u}_{\rm ref}(r;r^{'}) + {\rm u}_{\rm ac}(r;r^{'}) \label{eq:onephotonprocess} \\
&& {} \mbox{ with } {\rm u}_{\rm ref}(r;r^{'}) = -\delta \left(r^{'}-r\right) \nonumber \\
&& {} \mbox{ and } {\rm u}_{\rm ac}(r;r^{'}) = 
\begin{cases}
\frac{2\kappa}{c} \frac{-e^{-\frac{\Omega_{-}}{c}(r^{'}-r)}\Omega_{-}+ e^{-\frac{\Omega_{+}}{c}(r^{'}-r)}\Omega_{+}}{\Omega_{+} - \Omega_{-}} \\ \mbox{\ \ \ \ for } r < r^{'}\\
0 \mbox{\ \ for } r > r^{'}.
\end{cases} \nonumber
\end{eqnarray}
The component \({\rm u}_{\rm ref}\) is the reflected component for the reflected photon without being absorbed by the atom-cavity system, while \({\rm u}_{\rm ac}\) is the transition component for photons re-emitted by the atom-cavity system.

We can now calculate the fidelities $F^{\rm g}_{int}(d)$ and $F^{\xi_{2}}_{int}(d)$ using the input and the output wavefunctions as
\begin{eqnarray}
&& {} F^{\rm g}_{int}(d) = -\int^{\infty}_{-\infty} dr \ \psi_{in}(r+d) \cdot \psi_{out}(r) \mbox{ for } g_{1} > 0 \nonumber \\
&& {} \label{eq:fidelityg}\\
&& {} F^{\xi_{2}}_{int}(d) =\int^{\infty}_{-\infty} dr \ \psi_{in}(r+d) \cdot \psi_{out}(r) \mbox{ for } g_{1} = 0, \nonumber \\
\label{eq:fidelitye}
\end{eqnarray}
When $F^{\rm g}_{int}(d)$ is equal to $F^{\xi_{2}}_{int}(d)$, we denote the relative delay time $d$ by $d_{p}$. In eqs.~(\ref{eq:fidelityg}) and (\ref{eq:fidelitye}), the velocity of light $c$ is taken to be $1$.

\begin{figure}[htbp]
\begin{center} 
\includegraphics[width=6.5cm]{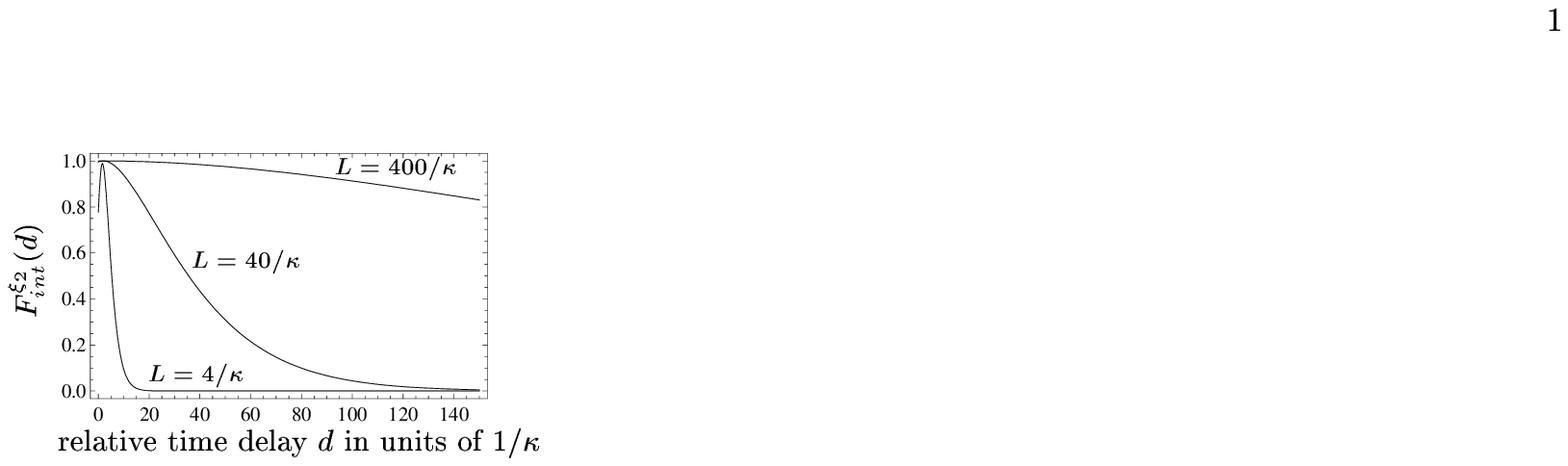}\\
\caption{\label{fig:cavityeffects} \scriptsize Fidelity $F^{\xi_{2}}_{int}(d)$ for an input pulse duration of $L = 4/\kappa$, $40/\kappa$, and $400/\kappa$.}
\end{center} 
\end{figure}

Figure~\ref{fig:cavityeffects} shows the fidelity $F^{\xi_{2}}_{int}(d)$ for the input wavefunction $\psi_{in}(r)=e^{- 2 \left|r\right|/L}/\sqrt{L/2}$. For a pulse duration of $L=4/\kappa$, $F^{\xi_{2}}_{int}(d)$ has a peak at the delay time $d \simeq 1/\kappa$ due to the relative pulse delay caused by the cavity decay rate $2\kappa$. As the pulse duration $L$ is increased, this effect becomes weak, as shown by the plots for $L=40/\kappa$ and $L=400/\kappa$ in Fig.~\ref{fig:cavityeffects}.

\begin{figure}[htbp]
\begin{center} 
\includegraphics[width=6.5cm]{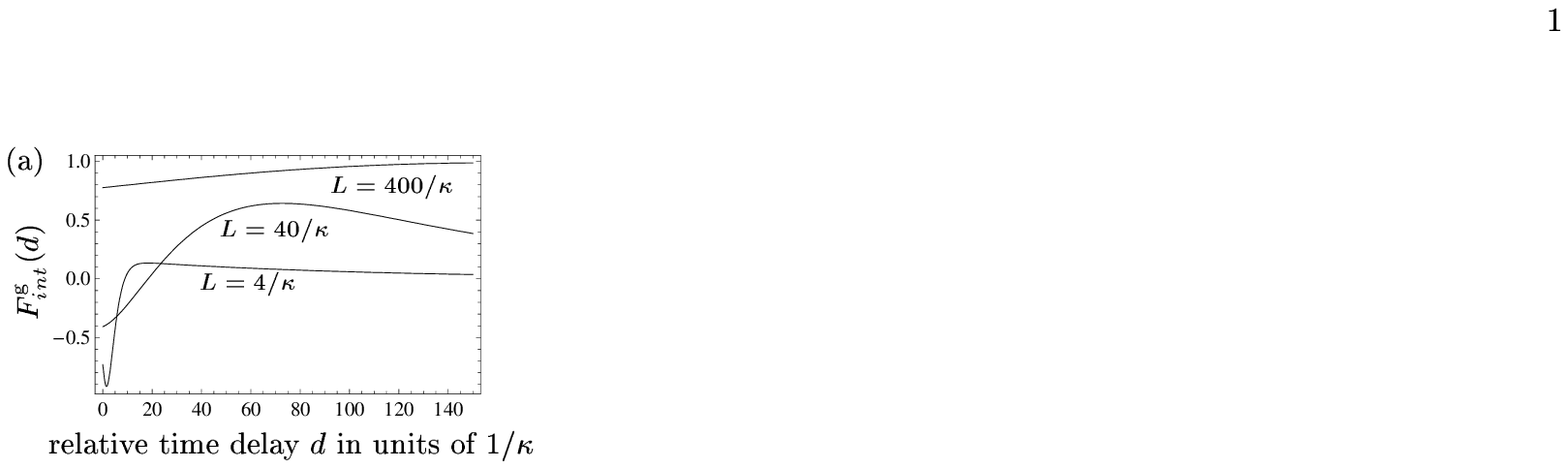}\\
\includegraphics[width=6.5cm]{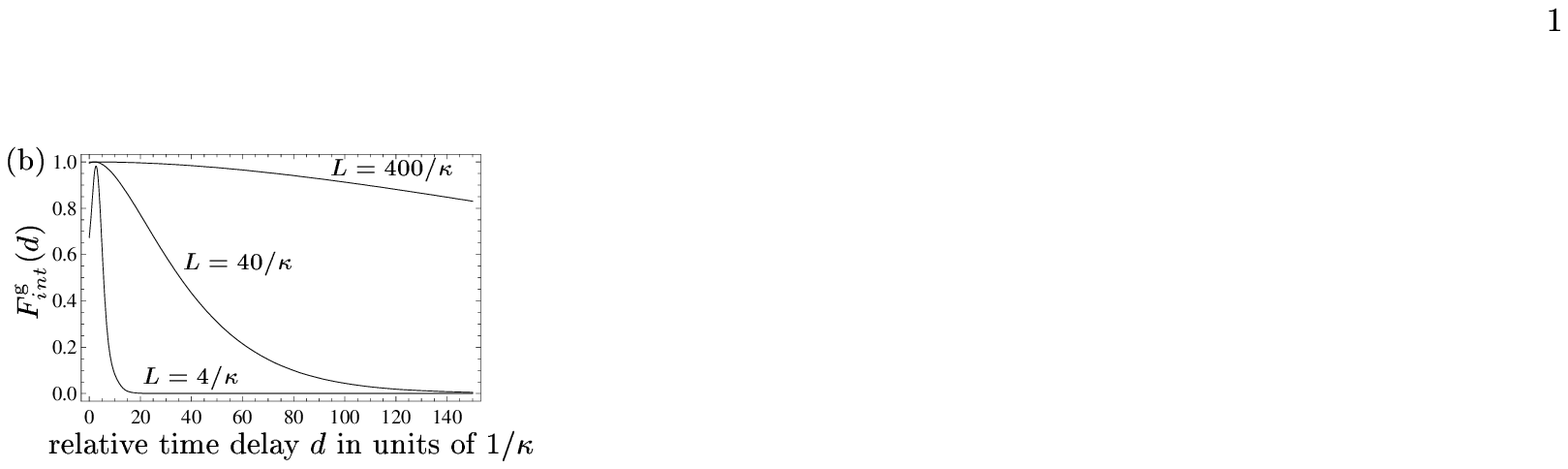}\\
\includegraphics[width=6.5cm]{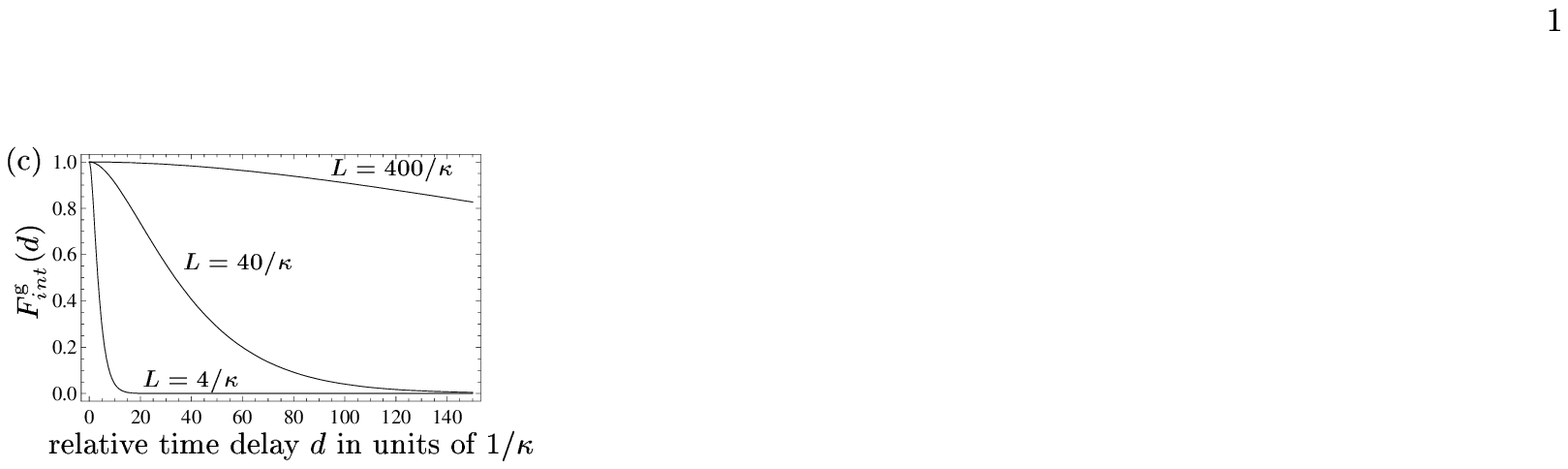}
\caption{\label{fig:couplingeffects} \scriptsize Fidelity $F^{\rm g}_{int}(d)$ for the input pulse duration $L = 4/\kappa$, $40/\kappa$, and $400/\kappa$. (a) the ratio $g_{1}/\kappa=0.1$, (b) $=1$, and (c) $=10$}
\end{center} 
\end{figure}

Figure~\ref{fig:couplingeffects}(a) shows the fidelity $F^{\rm g}_{int}(d)$ for a coupling constant of $g_{1} = 0.1 \kappa$. In such a leaky-cavity regime, the atomic system and the cavity mode do not interact cooperatively with the input pulse and there is no vacuum Rabi splitting. In this situation, the intracavity atomic system interacts with the input pulse through the single mode of the cavity. The relative $\pi$ phase flip from the input wavefunction is thus caused by the absorption and reemission processes of the intracavity atomic system \cite{kojima03b,kojima07}. The dipole relaxation rate of that atomic system is approximately $\Gamma_{1} \simeq g^{2}_{1}/\kappa=10^{-2}\kappa$. This implies that a high fidelity of $F^{\rm g}_{int}(d)$ can be obtained for input pulse durations larger than $100/\kappa$. Thus, $F^{\rm g}_{int}(d)$ for $L=4/\kappa$ and $L=40/\kappa$ do not exceed $0.1$ and $0.7$, respectively, as shown in Fig.~\ref{fig:couplingeffects}(a). The peak of $F^{\rm g}_{int}(d)$ for $L=40/\kappa$ is caused by the increase in the relative pulse delay due to the long dipole relaxation time, which is given approximately by $1/\Gamma_{1}=100/\kappa$. The gate fidelity ${\rm F}_{int}=F^{\rm g}_{int}(d_{p})=F_{int}^{\xi_{2}}(d_{p})$ for $=400/\kappa$ is $0.936$.  It should be noted that when the phase flip operation fails, the fidelity $F^{\rm g}_{int}(d)$ is negative. It has been shown that there is a small relative delay of less than the relative time delay $d=10/\kappa$ for $L=4/\kappa$ and $L=40/\kappa$ in Fig.~6(a). Indeed, this regime is not appropriate for CPF, because the effects of spontaneous emission with a rate of $2\gamma_{1}$ are not negligible unless the dipole relaxation rate $\Gamma_{1}$ is much larger than $\gamma_{1}$, which was not assumed in our model. In the following, this regime is not discussed for CPF.

Figure~\ref{fig:couplingeffects}(b) shows the fidelity $F^{\rm g}_{int}(d)$ for an input wavefunction and a coupling constant of $g_{1}=\kappa$. For a pulse duration of $L=4/\kappa$, the $F^{\rm g}_{int}(d)$ has a peak at the delay time $d=3/\kappa$ due to the relative pulse delay caused by the cavity decay rate $\kappa$. When the pulse duration $L$ is increased, this effect becomes weak, as shown by the plots for $L=40/\kappa$ and $L=400/\kappa$ in Fig.~\ref{fig:couplingeffects}(b). The vacuum Rabi splitting in the regime $g_{1}=\kappa$ suppresses absorption of the input photon pulse by the atom-cavity system and the reflection of that pulse by the cavity mirror becomes significant as the input pulse duration is increased. This reflection changes the phase of the input pulse by $\pi$ and therefore the maximum values of $F^{\rm g}_{int}(d)$ are $0.982$ for $L=4/\kappa$, $0.999$ for $=40/\kappa$, and $0.999$ for $=400/\kappa$. The gate fidelity was $0.964$ for $L=4/\kappa$, $0.999$ for $=40/\kappa$, and $0.999$ for $=400/\kappa$. We truncate numbers to three decimal places, because it is sufficient to discuss the performance of the PAD with an imperfect detector having a quantum efficiency of less than $10^{-1}$. More accurate calculations will need to be performed if we apply CPF to a C-NOT when quantum information processing with several qubits.

Figure~\ref{fig:couplingeffects}(c) shows the fidelity $F^{\rm g}_{int}(d)$ for the coupling constant $g_{1}=10\kappa$. The qualitative features are similar to the case for $g_{1}=\kappa$. The vacuum Rabi splitting in the regime $g_{1}=10\kappa$ strongly suppresses absorption of the input photon pulse by the atom-cavity system relative to the previous case. The maximum value of $F^{\rm g}_{int}(d)$ therefore increased to $0.999$ even when $L=4/\kappa$. However, the gate fidelity was reduced to 0.937 due to the effect of the relative pulse delay on the fidelity $F^{\xi_{2}}_{int}(d)$. The gate fidelities ${\rm F}_{int}$ for $L=40/\kappa$ and $=400/\kappa$ were $0.998$, and $0.999$, respectively.

The time for the interaction between the signal photon and the atom-cavity system for CPF needs to be much shorter than the dipole relaxation time $1/\Gamma_{2}$ of the excited state $\xi_{2}$ to avoid an unexpected transition to the excited state $\xi_{1}$. The interaction time $\delta t_{int}$ can be defined using the input pulse duration $L$ and the relative time delay $d_{p}$ for the gate fidelity ${\rm F}_{int}$. The input wavefunction has a probability of $0.999$ in the region $-1.75L \le r \le 1.75L$. The interaction time $\delta t_{int}$ should thus be $3.5L + d_{p}$. The dipole relaxation time $1/\Gamma_{2}$ has to be more than $10^{2} \cdot \delta t_{int}$ for the relaxation of the projection probability ${\rm P}_{h}(t)$ (given in eq.~(\ref{eq:halfrepresentation})) to be less than 1 \%. This means that the coupling constant $g_{2}$ should be less than $\sqrt{\kappa/(10^{2} \delta t_{int})}$. In the coupling regime $g_{1}=\kappa$ and $g_{1}=10\kappa$, the coupling constant $g_{2}$ should be less than $\kappa/40$ for $L=4/\kappa$, $\kappa/120$ for $40/\kappa$, and $\kappa/380$ for $400/\kappa$. That is, the ratio between the two oscillator strengths of the V-type three-level system: $(g_{1}/g_{2})^{2}$ should be more than $40^{2}$ in the strong coupling regime \(g_{1} \ge \kappa\). In particular, the regime \(g_{1} \simeq \kappa\) is better than the regime \(g_{1} \gg \kappa\) since it can achieve a gate fidelity of ${\rm F}_{int}=0.999$ for a short time duration of $L=40/\kappa$, and it minimizes the time taken for the interaction with the atom-cavity by increasing $g_{1}$. Furthermore, minimization of the interaction time minimizes the effects of spontaneous emission from the excited state $\ket{\xi_{2}}$ on the projection probability to the half-way state.

\section{Accuracy and efficiency of photon-arrival detector}

In the following discussion, the gate fidelity ${\rm F}_{int}$ is taken to be $0.999$ based on the results of the previous section which showed that this value suppresses the influence of entanglement between the signal photon and the intracavity atomic system on the atomic state after the second rotation ${\rm R}(-\pi/2)$ in the procedure of the PAD. In order to characterize the performance of PAD shown in Fig.~\ref{fig:qndscheme}, we introduce conditional probabilities ${\rm P}^{\rm PAD}_{ij}$ ($i,j=0,1$). An index $i=1 (0)$ means that there is a (no) signal photon in the signal input port and an index $j=1 (0)$ means that the detector Det1 has (not) detected a photon. 
We proceed to evaluate the conditional probabilities ${\rm P}^{\rm PAD}_{11}$ and ${\rm P}^{\rm PAD}_{01}$. The order of operation by the laser pulses is (i) $\pi/2$ rotation and (ii) $-\pi/2$ rotation. The rotation (i) converts the ground state into a half-way state. When there is a signal photon in the signal input port, the half-way state is changed into the orthogonal half-way state by the phase flip of $\pi$ in the ground state $\ket{g}$. Rotation (ii) then changes the orthogonal half-way state into the excited state $\ket{\xi_{2}}$. On the other hand, when there is no photon in the signal input port, the half-way state by rotation (i) is changed into the ground state $\ket{\rm g}$ by rotation (ii). We write the probability of the phase flip of $\pi$ as ${\rm P}_{\rm (i)}(t_{r} + \delta t_{int})$, where $\delta t_{int}$ corresponds to the interaction time of the signal photon with the atom-cavity system (see Sec.~V) after a time $t_{r}$ taken for rotation by the laser pulse. Likewise, the projection probabilities of the density matrix of the atomic system $\rho_{atom}(t)$ to the excited state starting from the half-way state, and to the ground state starting from the orthogonal half-way state are written as ${\rm P}^{\xi_{2}}_{\rm (ii)}(t_{r} + \tau)$, and ${\rm P}^{\rm g}_{\rm (ii)}(t_{r} + \tau)$ respectively, where the delay time $\tau$ corresponds to the start time for the detection of a Y-polarized photon at {\rm Det1} after rotation (ii).

The temporal evolution of the intracavity atomic system during the interaction time $\delta t_{int}$ is dominated by the dipole relaxation of the excited state $\ket{\xi_{2}}$ through the cavity mode, and by the interaction with the signal photon. The generator of the temporal evolution is given by $\hat{H}^{(1)}-i\Gamma_{2} \ketbra{\xi_{2}}{\xi_{2}}$, where the first term is the Hamiltonian for the interaction of the atom-cavity system with the signal photon $\hat{H}^{(1)} = \hat{H}^{(1)}_{F_{c}}+\hat{H}^{(1)}_{int F_{c}}+ \hat{H}^{(1)}_{int ac}$ given in eq.~(\ref{eq:exthamiltonian}) and the second term effectively describes the generator of the dipole relaxation by the rate $\Gamma_{2}$ in the intracavity atomic system. This is because, the equations of motion for the longitudinal and transversal components, $\bracket{\sigma^{(2)}_{-}(t)}$ and $\bracket{\sigma^{(2)}_{3}(t)}$, derived from the Heisenberg equation of motion with the total Hamiltonian $\hat{H}$ given by eq.~(\ref{eq:exthamiltonian}) are the same as those derived from the generator $\hat{H}^{(1)}-i\Gamma_{2} \ketbra{\xi_{2}}{\xi_{2}}$ for the input field amplitude $\epsilon_{in} = 0$ in the leaky-cavity regime. Since the relaxation term $-i\Gamma_{2} \ketbra{\xi_{2}}{\xi_{2}}$ commutes with the Hamiltonian $\hat{H}^{(1)}$, the atomic density matrix during the interaction time $\delta t_{int}$ is given by
\begin{eqnarray}
{\rm Tr}_{1ph} \left[ \exp \left[-(i/\hbar) \hat{H}^{(1)} t \right] \rho_{1ph} \otimes \rho_{atom}(t) \exp \left[(i/\hbar) \hat{H}^{(1)}t \right] \right]  \nonumber \\
\label{eq:onephtrout}
\end{eqnarray}
, where $\rho_{atom}(t)$ is the same as eq.~(\ref{eq:halfrepresentation}) for the input field amplitude $\epsilon_{in}=0$, and ${\rm Tr}_{1ph}\left[\ldots\right]$ traces the signal photon state. Thus, the probability of phase flip $\pi$: ${\rm P}_{\rm (i)}(t_{r} + \delta t_{int})$ corresponds to the probability ${\rm P}_{h}(t_{r} + \delta t_{int}) \times {\rm F}_{int}$ using eq.~(\ref{eq:onephtrout}) and eq.(\ref{eq:halfrepresentation}) starting from the ground state $\ket{\rm g}$. Likewise, the probabilities ${\rm P}^{\xi_{2}}_{\rm (ii)}(t_{r} + \tau)$ and ${\rm P}^{\rm g}_{\rm (ii)}(t_{r} + \tau)$ correspond to the probabilities $(1 + P_{3}(t_{r}+\tau))/2$ and $(1-P_{3}(t_{r}+\tau))/2$, respectively. The conditional probability ${\rm P}^{\rm PAD}_{11}$ can thus be described as
\begin{eqnarray}
{\rm P}^{\rm PAD}_{11} = && {} {\rm P}_{\rm (i)}(t_{r} + \delta t_{int}) \cdot {\rm P}^{\xi_{2}}_{\rm (ii)}(t_{r} + \tau) \cdot {\rm P}_{\rm eff} \nonumber \\
&& {} + {\rm P}_{\rm noise} (\tau), \label{eq:qndsuccess}
\end{eqnarray}
where the probabilities ${\rm P}_{\rm eff}$ and ${\rm P}_{\rm noise} (\tau)$ correspond to the quantum efficiency of the detector {\rm Det1} and the average intracavity photon number after the delay time $\tau$, respectively.
Likewise,
\begin{eqnarray}
{\rm P}^{\rm PAD}_{01} = && {} \left(1-{\rm P}_{h}(t_{r} + \delta t_{int})\right) \nonumber \\
&& {} \times {\rm P}^{\xi_{2}}_{\rm (ii)}(t_{r} + \tau) \cdot {\rm P}_{\rm eff} + {\rm P}_{\rm dark} + {\rm P}_{\rm noise} (\tau), \nonumber \\
\label{eq:qndunsuccess}
\end{eqnarray}
where the probability $\left(1-{\rm P}_{h}(t_{r} + \delta t_{int})\right)$ corresponds to that for the failure of the initial rotation to produce a desired half-way state. The state corresponding to that failure is the other half-way state orthogonal to the desired half-way state. As mentioned in Sec.~III, such an orthogonal state causes the emission of probe photons when there is no photon in the signal input port. The probability ${\rm P}_{\rm dark}$ corresponds to the dark count rate of the detector {\rm Det1}.

In the regime $\kappa=g_{1}$ and $g_{2}=\kappa/120$, we calculated the conditional probabilities ${\rm P}^{\rm PAD}_{11}$ and ${\rm P}^{\rm PAD}_{01}$. When calculating the projection probabilities ${\rm P}_{\rm (i)}(t_{r} + \delta t_{int})$ and ${\rm P}^{\xi_{2}}_{\rm (ii)}(t_{r} + \tau)$, we adapted the input pulse duration $L=40/\kappa$ and the average input photon number $\bar{n}_{in}=10^{4}$. Using eqs.~(\ref{eq:bloch}) and (\ref{eq:halfrepresentation}), we get ${\rm P}_{\rm (i)}(t_{r} + \delta t_{int})$ and ${\rm P}^{\xi_{2}}_{\rm (ii)}(t_{r} + \tau)$ are $0.994$ and $0.981$ at $t_{r}=80.64/\kappa$, $\delta t_{int} = 144/\kappa$ and $\tau = 4.2/\kappa$, respectively. The conditional probability ${\rm P}^{\rm PAD}_{11}$ is thus $0.974 \cdot 10^{-1} + 4.5 \cdot 10^{-4}=9.785 \cdot 10^{-2}$, where the second term corresponds to the probability ${\rm P}_{\rm noise} (\tau)$. The probability ${\rm P}_{\rm noise}$ was initially $\bar{n}_{a}=200$ for $\bar{n}_{in}=10^{4}$ and then decreased to ${\rm P}_{\rm noise}(\tau)=\bar{n}_{a} \cdot \exp [-\kappa \tau] = 4.5 \cdot 10^{-4}$ after the delay time $\tau=4.2/\kappa$. Likewise, the conditional probability ${\rm P}^{\rm PAD}_{01}$ is $5.88 \cdot 10^{-3} \cdot 10^{-1} + 10^{-5} + 4.5 \cdot 10^{-4}=10.48 \cdot 10^{-4}$, where the second term corresponds to the dark count rate of Det1, which was assumed to be ${\rm P}_{\rm dark} = 10^{-5}$ in units of $10/\Gamma_{2}=(1.6 \cdot 10^{4})/\kappa$.

The proposed PAD allows us to apply the procedure of the PAD many times to detect photon arrival. In the following, we discuss the accuracy of the detection of photon arrival after repeated applications. That accuracy can be evaluated by the average number of counts at Det1 due to photon arrival divided by the average total number of counts at Det1. Describing the linear transmittance to return a signal photon to the signal input port as ${\rm P}_{\rm T}$ per round trip, the average total number of counts at Det1 after $n$ applications in the PAD can be written as
\begin{eqnarray}
&& {} \bar{\rm N}^{\rm Det1} = {\rm P}_{1} \bar{\rm N}^{\rm Det1}_{1} + {\rm P}_{0} \bar{\rm N}^{\rm Det1}_{0} \nonumber \\
&& {} \mbox{, where } \bar{\rm N}^{\rm Det1}_{1} = {\rm P}^{\rm PAD}_{11} \cdot \sum^{n}_{k=1} ({\rm P}_{\rm T})^{k-1} \nonumber \\
&& {} \ \ \ \ \ \ \ \ \ \ \ \ \ \ \ \ \ \ \ \ \ \ + {\rm P}^{\rm PAD}_{01} \cdot \left(n-\sum^{n}_{k=1} ({\rm P}_{\rm T})^{k-1}\right) \nonumber \\
&& {} \mbox{, and }  \bar{\rm N}^{\rm Det1}_{0} = n \cdot {\rm P}^{\rm PAD}_{01} \nonumber \\
&& {}
\end{eqnarray}
, where ${\rm P}_{1}$ is the probability that there is a signal photon in the signal input port before starting the PAD procedure, and ${\rm P}_{0}$ is the probability that there is no photon in the input port. $\sum^{n}_{k=1} ({\rm P}_{\rm T})^{k-1}$ in the equation above corresponds to the average number of times that a signal photon appears in the signal input port in $n$ attempts.
 
The accuracy of detecting photon arrivals after the repeated applications can thus be described by in terms of the average number of counts at Det1 due to photon arrivals, ${\rm P}_{1} \cdot {\rm P}^{\rm PAD}_{11} \cdot \left(\sum^{n}_{k=1} ({\rm P}_{\rm T})^{k-1} \right)$,
\begin{eqnarray}
&& {} {\rm P}_{\rm PAD} \equiv {\rm P}_{1} \left({\rm P}^{\rm PAD}_{11} \cdot \sum^{n}_{k=1} ({\rm P}_{\rm T})^{k-1}\right)/\bar{\rm N}^{\rm Det1} \nonumber \\
\end{eqnarray}
for $n$ applications. 

Referring to the practical use of the repeated applications, once a count is registered by detector Det1 in Fig. 2, that application is stopped and the next signal input is processed in the same way. The signal input state characterized by probabilities ${\rm P}_{0}$ and ${\rm P}_{1}$ is finally projected into the state with a single count at Det1. ${\rm P}_{\rm PAD}$ corresponds to the ratio of a single photon in the state projected by one count at $\rm Det1$. If the probability ${\rm P}_{\rm PAD}$ is equal to $1$, repeated applications of the PAD projects a signal input state into a single photon state with the detection of photon arrival by counts at the detector $\rm Det1$.

\begin{figure}[htbp]

\begin{center} 
\includegraphics[width=6.5cm]{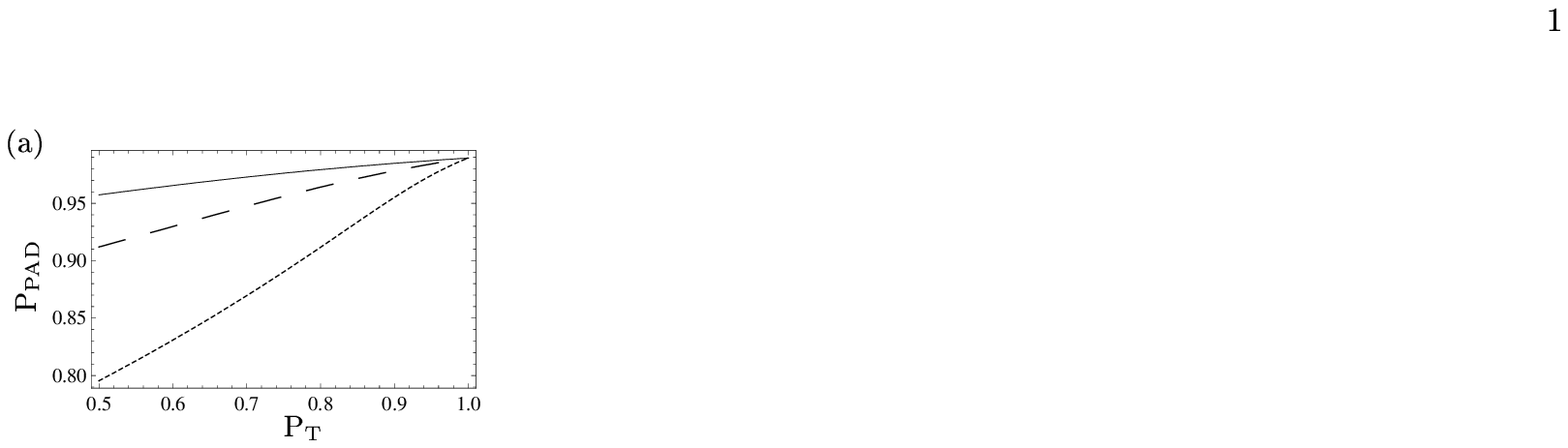}\\
\includegraphics[width=6.5cm]{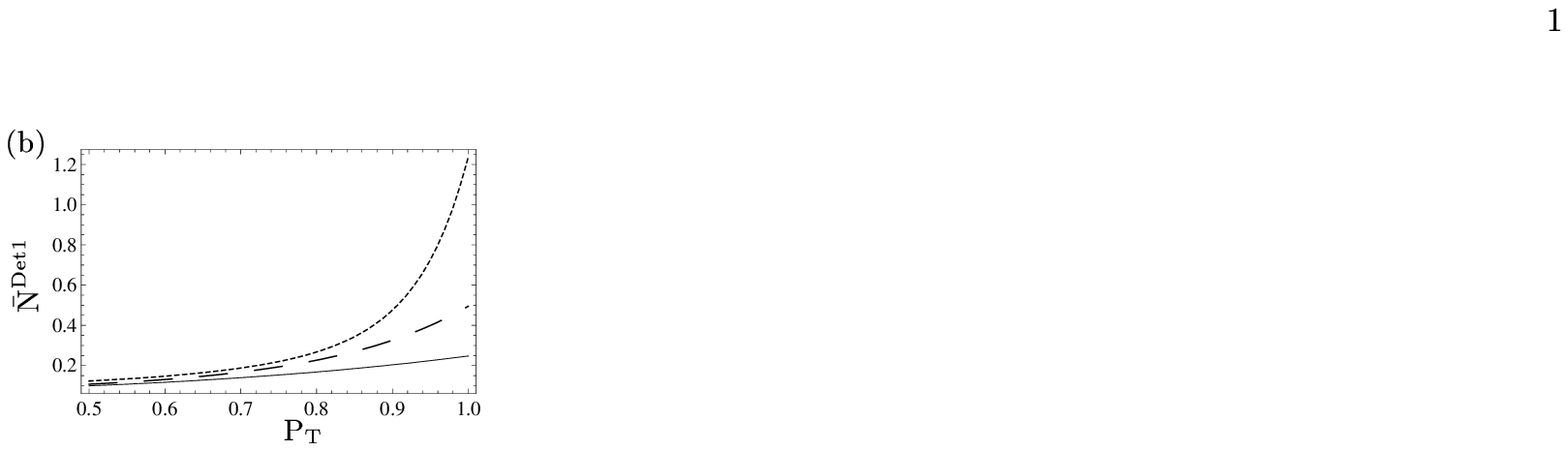}
\caption{\label{fig:qndsuccess} \scriptsize Success probabilities and average detection numbers for $n=5$ (solid line), $10$ (broken line), and $25$ (dotted line) attempts for ${\rm P}_{0}={\rm P}_{1}=1/2$, ${\rm P}^{\rm PAD}_{11}=0.0974 + 4.5 \cdot 10^{-4}=9.785 \cdot 10^{-2}$, and ${\rm P}^{\rm PAD}_{01}=5.88 \cdot 10^{-3} \cdot 10^{-1} + 10^{-5}+4.5 \cdot 10^{-4}=10.48\cdot10^{-4}$. Dependence of (a) ${\rm P}_{\rm PAD}$ and (b) $\bar{\rm N}^{\rm Det1}$ on linear transmittance ${\rm P}_{\rm T}$.}
\end{center} 
\end{figure}

Figure~\ref{fig:qndsuccess}(a) shows the dependence of the accuracy of detecting photon arrival ${\rm P}_{\rm PAD}$ on the linear transmittance ${\rm P}_{\rm T}$ after $n=5$ (solid line), $10$ (broken line), and $25$ (dotted line) applications for a random input ${\rm P}_{0}={\rm P}_{1}=1/2$, ${\rm P}^{\rm PAD}_{11}=9.785 \cdot 10^{-2}$, and ${\rm P}^{\rm PAD}_{01}=10.48\cdot10^{-4}$. ${\rm P}_{\rm PAD}$ increases with increasing linear transmittance ${\rm P}_{\rm T}$ and decreases with increasing number of applications. The corresponding average number of counts at Det1 increases with increasing linear transmittance, and with increasing number of applications. We evaluate the efficiency of the proposed PAD by comparing with the responses of ideal photodetectors. Since ideal photodetectors detect signal photon by probability ${\rm P}_{1}$, we define the efficiency of the PAD (${\rm P}^{*}_{\rm PAD}$) with the accuracy of detecting photon arrival ${\rm P}_{\rm PAD}$ when the average total number of counts $\bar{\rm N}^{\rm Det1}$ is equal to the probability ${\rm P}_{1}$. If the linear transmittance ${\rm P}_{\rm T}$ is larger than $0.9$, the efficiency of the PAD exceeds $0.95$ for $n \le 25$, as shown in Fig. 7(a).

\section{Discussion of experimental realization}
So far, we have neglected the effects of spontaneous emission on $\gamma_{2}=\gamma_{1}g_{2}/g_{1}$. However, since $\gamma_{2}$ can be comparable to the radiative relaxation rate $\Gamma_{2}$, spontaneous emission may affect the projection probabilities ${\rm P}_{\rm (i)}$ and ${\rm P}^{\xi_{2}}_{\rm (ii)}$ given in eqs.~(\ref{eq:qndsuccess}) and (\ref{eq:qndunsuccess}). For example, $\gamma_{2}$ is approximately equal to $1.2\Gamma_{2}$ for $g_{1}/g_{2}=120$ and $\gamma_{1}=10^{-2} \kappa$. In this case, the probabilities ${\rm P}^{\rm PAD}_{11}$ and ${\rm P}^{\rm PAD}_{01}$ calculated in the previous section, become $0.0947 + 4.5 \cdot 10^{-4}=9.515 \cdot 10^{-2}$ and $11.75 \cdot 10^{-3} \cdot 10^{-1} + 10^{-5} + 4.5 \cdot 10^{-4}=16.35 \cdot 10^{-4}$, respectively, due to the decrease in the probabilities ${\rm P}_{\rm (i)}$ and ${\rm P}^{\xi_{2}}_{\rm (ii)}$. The resultant accuracy ${\rm P}_{\rm PAD}$ is then $0.983$ at the limit ${\rm P}_{\rm T}=1$. The performance of the PAD is about 0.006 less than that when there is no noise due to spontaneous emission. This result means that spontaneous emission with a rate $2\gamma_{2}$ will not be a source of error in the strong coupling regime $g_{1} \ge \kappa$, if the spontaneous emission rate $2\gamma_{1}$ is much smaller than the cavity decay rate $2\kappa$.

In Sec.~V, we demonstrated that a ratio between the two oscillator strengths of the V-type three-level system of \(40^{2} \le (g_{1}/g_{2})^{2}\) is required for a high gate fidelity of the quantum phase flip operation. Such a high ratio may be quite difficult to realize for real atoms. However, artificial molecules consisting of semiconductor nanostructures are a potential solution. Note that the oscillator strength is proportional to the convolution between the electron wavefunction and the hole wavefunction in a neutral exciton. Therefore, increasing the average distance between an electron and a hole will cause the corresponding oscillator strength to decrease. A single quantum dot molecule (e.~g.,~coupled disk quantum dots) allows us to engineer the ratio between the oscillator strengths of the intradot exciton and the interdot exciton to be extremely asymmetric by controlling the distance between two dots \cite{tabic}. Different polarization selection rules for intradot and interdot excitons can be achieved by g-factor engineering\cite{kosaka}. A promising candidate for an atom-cavity system for the experimental realization of the PAD will thus be a quantum-dot molecule in a confocal or photonic crystal cavity (or both).

Experimentally, the dephasing time of the superposition of the ground and the excited states is much faster than the lifetime of the excited state and it will be less than 1 ns \cite{nakaoka} due to, for example, electron-electron scattering with residual electrons outside the quantum dots at a low temperature of a few K \cite{fasol}. Thus, the pulse time of the input photons has to be much less (much less than 1 ns) than the dephasing time to ensure that the half-way atomic state interacts with the photons. The cavity decay rate \(\kappa\) should be larger than a few hundred gigahertzs (\(200 \sim 300\) GHz) in order to excite a X-polarized cavity mode with such a short pulse. From the results of the fidelity analysis of the phase flip operation given in Sec.~V, the coupling strength \(g_{1}\) has to be comparable to, or larger than the cavity decay rate to achieve a high gate fidelity by the interaction of photon pulses with the atom-cavity system in a few tens of picoseconds, which is much shorter than the dephasing time.

To achieve such a large coupling strength, a small cavity with a volume comparable to the third power of the transition wavelength, is necessary, since the coupling strength is proportional to the dipole moment of the transition to the excited state \(\xi_{1}\) and the square root of the cavity mode volume. Note that the dipole moment of light-hole exciton in a GaAs quantum dot is typically 20 D. Confocal cavities consisting of two curved mirrors can theoretically achieve a maximum coupling strength \(g_{1}\) of more than \(100\) GHz with a radius of curvature of less than 2 \(\mu\){\rm m} (calculated assuming a refractive index of 3.4, a cavity length of 1 \(\mu\)m, and a transition frequency of \(2 \times 10^{14}\) Hz and using the theories described in \cite{reitzen, yariv}). Another potential solution is 2D-photonic-crystal cavities since they can achieve a maximum coupling strength \(g_{1}\) of larger than 200 GHz by realizing cavity mode volumes of less than the third power of the transition wavelength in the range of \(\lambda = 0.9 \backsim 1.5\) \(\mu\){\rm m} \cite{shirane}.

A remaining problem is the influence of dephasing during Rabi rotation by laser pulses. As discussed in Sec.~IV, when the \(g_{2}\) is assumed to be \(10^{-2} \cdot \kappa\), the average photon number \(\bar{n}_{in}=10^{4}\) satisfies the leaky-cavity regime. In this case, Rabi rotation to the half-way state can be achieved in a few tens of picoseconds, which is much shorter than the dephasing time.

In summary of this section, we require that the cavity decay rate and the coupling strength of the transition to the excited state \(\xi_{1}\) is more than one hundred times greater than the dephasing rate between the ground state {\rm g} and the excited state \(\xi_{2}\). This requirement can be satisfied using semiconductor quantum dots and cavities. Furthermore, using a high decay rate and a coupling strength of a few hundred gigahertz will enable us to perform PAD and C-NOT operations at a high repetition rate.
\section{Comparison of repetition rates of Duan and Kimble scheme and our scheme}
It is interesting compare the repetition rates of our scheme with those of Duan and Kimble's scheme \cite{duan}. The difference between our scheme and their scheme is the atomic internal state used for CPF. In our scheme, the rotational operation between the ground state $\ket{\rm g}$ and one of the two excited states $\ket{\xi_{2}}$ is performed using laser pulses, through the cavity mode coupled with the excited state. On the other hand, in the Duan-Kimble scheme, the rotational operation should be conducted between two ground states, although the method for achieving this was not specified in their paper \cite{duan}. If Rabi oscillation by laser pulses is used, rotation is achieved by using two laser pulses having orthogonal polarizations, as the result of quantum interference between the two half-way states. Note that the rotation in their scheme cannot be performed through cavity modes within the framework of the $\lambda$-type system. This is because one of the two ground states is coupled with the cavity mode in a strong coupling regime rather than the leaky cavity regime. We therefore assume that rotational operations are conducted by directly irradiating laser pulses through the gap between cavity mirrors. The transverse dimension (Gaussian waist) of a confocal cavity consisting of two mirrors that have equal radii of curvature can be calculated by simple Gaussian beam propagation theory, which gives a good approximation when the radius of curvature of the mirrors is much larger than the cavity length (although it is not strictly accurate for length scales less than a few times the cavity length). The minimum cavity length of a confocal cavity (in which the laser beam passes through the gap between the edges of the confocal mirrors) should be larger than $15 \cdot \lambda_{a}$ to ensure that the average scattered photon number is less than $10^{-2}$. The corresponding cavity mode volume for a radius of curvature of $10^{5} \cdot \lambda_{a}$ is about $1.3 \cdot 10^{4} \cdot \lambda_{a}^{3}$, where $\lambda_{a}$ is the transition wavelength of the intracavity atomic system. On the other hand, the minimum cavity length of a cavity that does not have the laser access is $\lambda_{a}/2$. The minimum mode volume is about $78.4 \cdot \lambda_{a}^{3}$. Thus, the ratio of the mode volume without laser access to that with the laser access is thus $165.65$.  The atom-photon coupling rate is inversely proportional to the square root of the mode volume. The atom-photon coupling rate $g_{1}$ for the case without laser access is thus 12.87 times larger than that for the case with laser access. This means that our scheme can achieve atom-photon CPF $12.87$ times faster than the Duan and Kimble scheme with a $\lambda$-type system for the regime $g_{1} \ge \kappa$, since we assume that their scheme does not perform rotations between two ground states through the cavity mode by laser pulses.

Indeed, the performance of our scheme for the atom-photon CPF on applications is not determined solely by the results given above since the time taken for rotation generated by laser pulses is not negligible. Since the Duan-Kimble scheme for CPF was proposed with the intention of applying it to two-photon C-NOT, we compare our scheme with their scheme for application to C-NOT. The Duan-Kimble scheme for two-photon C-NOT consists of three atom-photon CPF and two rotational operations. The total operation time for C-NOT is $3 \cdot \delta t^{'}_{int}$ for $\lambda$-type atomic systems since the time taken for those rotations is negligible. $\delta t^{'}_{int}$ is the time taken for CPF. On the other hand, the time taken for those rotations is not negligible for our scheme. The minimum time is about 10 times of the time taken for CPF $\delta t_{int}$ in our scheme. It is given by $10 \cdot \delta t^{'}_{int}/12.87$ by using the time taken for the CPF in their scheme. The total operation time for C-NOT is thus $3 \cdot \delta t^{'}_{int}/12.87 + 2 \cdot 10 \cdot \delta t^{'}_{int}/12.87=1.787 \cdot \delta t^{'}_{int}$. Our scheme in two-photon C-NOT will be $1.679$ times faster than that of the Duan-Kimble scheme due to the minimization of the cavity size. For a "dream" mirror with a radius of curvature of the order of the transition wavelength $\lambda_{a}$, it might be possible to achieve a C-NOT that is 6.5 times faster than the Duan-Kimble scheme. For circuits including more CPFs than rotational operations, the difference in the repetition rates will be significant.

\section{Conclusion}
Detection of photon arrival based on a CPF using a V-type system was discussed. The gate fidelity for the CPF was calculated by analyzing the responses of the atom-cavity system to single photon input. A maximum gate fidelity of more than 0.99 in the strong coupling regime $g = \kappa \gg \gamma_{1}$ was achieved. The efficiency of the proposed PAD was estimated by analyzing the responses of an atom in a one-sided cavity for laser pulse input and calculating the accuracy and the average total number of counts. An efficiency of up to $98.9 \%$ was found to be possible by increasing the transmittance per round trip of a signal photon.

There are two drawbacks in the present scheme of the PAD. One is that it takes a long time ($\simeq 10/\Gamma_{2}$) to initialize the intracavity atomic system to the ground state $\ket{\rm g}$ before starting the PAD procedure. A long initialization time decreases the repetition rate of the PAD procedure. The other drawback is that the accuracy ${\rm P}_{\rm PAD}$ is smaller than 0.989 due to the linear transmittance ${\rm P}_{\rm T}$ being smaller than 1. 
These problems may be solved by direct detection of the excited state $\ket{\xi_{2}}$ in a quantum dot molecule with a quantum point contact.

Nevertheless, the proposed PAD procedure enables us to apply the PAD procedure many times to detect the arrival of input photons and the average number of counts $\bar{\rm N}^{\rm Det1}_{1}$ exceeds the quantum efficiency of photodetectors in a small number (less than ten) of applications, when the linear transmittance is larger than the quantum efficiency. Furthermore, by minimizing the cavity size our scheme for atom-photon CPF may be capable of achieving a higher repetition rate than the Duan-Kimble scheme \cite{duan}.

\section*{Acknowledgement}
K.~K. thanks NEC researchers S.~Ishizaka, S.~Kono, M.~Shirane, A.~Kirihara, K.~Yoshino, AIST researchers H.~Imamura, Y.~Rikitake, and IIS of Tokyo Univ. Project Lecturer T.~Nakaoka for grateful discussions.


\begin{thebibliography}{99}
\bibitem{bennet} C.~H.~Bennet, G.~Brassard, C.~Crepeau, R.~Jozsa, A.~Peres, and W.~K.~Wootters, Phys.~Rev.~Lett {\bf 70}, 1895 (1993). 
\bibitem{duan} L.-M.~Duan and H.~J.~Kimble, Phys.~Rev.~Lett. {\bf 92} 127902 (2004).
\bibitem{turchette} Q.~A.~Turchette et.~al, Phys.~Rev.~Lett. {\bf 75}, 4710 (1995).
\bibitem{loock} P.~van Loock, T.~D.~Ladd, K.~Sanaka, F.~Yamaguchi, K. Nemoto, W.~J.~Munro, and Y. Yamamoto, Phys.~Rev.~Lett. {\bf 96} 240501 (2006).
\bibitem{kuhn} A.~Kuhn, M.~Hennrich, and G.~Rempe, Phys.~Rev.~Lett. {\bf 89} 067901 (2002). 
\bibitem{Wu} Y.~Wu, X.~Li, L.~M.~Duan, D.~G.~Steel, and D.~Gammon, Phys.~Rev.~Lett. {\bf 96} 087402 (2006).
\bibitem{kosaka-crest} H.~Kosaka, Y.~Mitsumori, Y.~Rikitake, and H.~Imamura, Appl.~Phys.~Lett. {\bf 90}, 113511 (2007); H.~Kosaka, H.~Shigyou, Y.~Mitsumori, Y.~Rikitake, H.~Imamura, T.~Kutsuwa, K.~Arai, and K.~Edamatsu, Phys.~Rev.~Lett. {\bf 100} 096602 (2008). \bibitem{tabic} M.~Tadic and F.~M.~Peeters, J. Phys.: Condens. Matter {\bf 16} 8633 (2004).
\bibitem{kosaka} H.~Kosaka, D.~S.~Rao, H.~D.~Robinson, P.~Bandaru, K.~Makita, and E.~Yablonovitch, Phys. Rev. B {\bf 67}, 045104 (2003).
\bibitem{multimode}K.~Kojima, H.~F.~Hofmann, S.~Takeuchi, and K.~Sasaki, Phys.~Rev.~A {\bf 70} 013810 (2004).
\bibitem{kojima07} K.~Kojima and A.~Tomita, Phys.~Rev.~A {\bf 75} 032320 (2007).
\bibitem{kojima03b}K.~Kojima, H.~F.~Hofmann, S.~Takeuchi, and K.~Sasaki, quant-ph/0404119. 
\bibitem{nakaoka} T.~Nakaoka, E.~C.~Clark, H.~J.~Krenner, M.~Sabathil, M.~Bichler, Y.~Arakawa, G.~Abstreiter, and J.~J.~Finley, Phys.~Rev.~B {\bf 74} 121305 (2006).
\bibitem{fasol} G.~fasol, Appl.~Phys.~Lett. {\bf 61}, 831 (1992).
\bibitem{yariv} A.~Yariv, Optical Electronics in Modern Communications, Fifth edition (Oxford Univ.~Press, 1997).
\bibitem{reitzen} S.~Reitzenstein, A.~L\"offler, C.~Hofmann, A.~Kubanek, M.~Kamp, J.~P.~Reithmaier, A.~Forchel, V.~D.~Kulakovskii, L.~V.~Keldysh, I.~V.~Ponomarev, and T.~L.~Reinecke, Opt.~Lett vol. 31, no. 11 1738 (2006).
\bibitem{shirane} M. Shirane, S. Kono, J. Ushida, S. Ohkouchi, N.~Ikeda, Y.~Sugimoto, and A. Tomita, J.~Appl.~Phys 101, 073107 (2007).

\end{thebibliography}
\end{document}